\documentclass[12pt,preprint]{aastex}
\newcommand{\bdv}[1]{\mbox{\boldmath$#1$}}

\def\au{{\rm au}} 
 
\def\kms{{\rm km}\,{\rm s}^{-1}}
\def\masyr{{\rm mas}\,{\rm yr}^{-1}}
\def\kpc{{\rm kpc}}
\def\mas{{\rm mas}}

\def\muas{\mu{\rm as}}

\def\pc{{\rm pc}}

\def\max{{\rm max}}
\def\min{{\rm min}}
\def\rel{{\rm rel}}

\def\eff{{\rm eff}}

\def\hel{{\rm hel}}

\def\fwhm{{\rm FWHM}}
\def\e{{\rm E}}
\def\bpi{{\bdv\pi}}
\def\bmu{{\bdv\mu}}

\def\btheta{{\bdv\theta}}

\def\bv{{\bf v}}

\begin{document}
\title{KMT-2022-BLG-2397: Brown Dwarf at the Upper Shore of the Einstein Desert}

\author{\textsc{
Andrew Gould$^{1,2}$
Yoon-Hyun Ryu$^{3}$, 
Jennifer C. Yee$^{4}$, 
Michael D. Albrow$^{5}$, 
Sun-Ju Chung$^{3,4}$, 
Cheongho Han$^{6}$, 
Kyu-Ha Hwang$^{3}$, 
Youn Kil Jung$^{3,7}$, 
In-Gu Shin$^{4}$, 
Yossi Shvartzvald$^{8}$, 
Hongjing Yang$^{9}$, 
Weicheng Zang$^{9}$,
Sang-Mok Cha$^{3,10}$, 
Dong-Jin Kim$^{3}$,
Seung-Lee Kim$^{3,7}$, 
Chung-Uk Lee$^{3}$,
Dong-Joo Lee$^{3}$,
Yongseok Lee$^{3,10}$, 
Byeong-Gon Park$^{3,7}$, 
Richard W. Pogge$^{2}$\\
(The KMTNet Collaboration)\\
}}
%----------------------------------------------------------------
\affil{$^{1}$Max-Planck-Institute for Astronomy, K\"{o}nigstuhl 17,
69117 Heidelberg, Germany}

\affil{$^{2}$Department of Astronomy, Ohio State University, 140 W.
18th Ave., Columbus, OH 43210, USA}

\affil{$^{3}$Korea Astronomy and Space Science Institute, Daejon
34055, Republic of Korea}

\affil{$^{4}$ Center for Astrophysics $|$ Harvard \& Smithsonian, 60 Garden
St., Cambridge, MA 02138, USA}

\affil{$^{5}$University of Canterbury, Department of Physics and
Astronomy, Private Bag 4800, Christchurch 8020, New Zealand}

\affil{$^{6}$Department of Physics, Chungbuk National University,
Cheongju 28644, Republic of Korea}

\affil{$^{7}$Korea University of Science and Technology, Korea, 
(UST), 217 Gajeong-ro, Yuseong-gu, Daejeon, 34113, Republic of Korea}

\affil{$^{8}$Department of Particle Physics and Astrophysics, 
Weizmann Institute of Science, Rehovot 76100, Israel}

\affil{$^{9}$ Department of Astronomy,
Tsinghua University, Beijing 100084, China}

\affil{$^{10}$School of Space Research, Kyung Hee University,
Yongin, Kyeonggi 17104, Republic of Korea}

\begin{abstract}

  We measure the Einstein radius of the single-lens microlensing event
  KMT-2022-BLG-2397 to be $\theta_\e=24.8\pm 3.6\,\muas$, placing it at the
  upper shore of the Einstein Desert, $9\la \theta_\e/\muas\la 25$,
  between free-floating planets (FFPs) and bulge brown dwarfs (BDs).
  In contrast to the six BD ($25\la\theta_\e\la 50$) events presented
  by \citet{gould22}, which all had giant-star source stars,
  KMT-2022-BLG-2397 has a dwarf-star source, with angular radius
  $\theta_*\sim 0.9\,\muas$.  This prompts us to study the relative
  utility of dwarf and giant sources for characterizing FFPs and BDs
  from finite-source point-lens (FSPL) microlensing events.  We find
  ``dwarfs'' (including main-sequence stars and subgiants) are likely
  to yield twice as many $\theta_\e$ measurements for BDs and a
  comparable (but more difficult to quantify) improvement for FFPs.
  We show that neither current nor planned experiments will yield
  complete mass measurements of isolated bulge BDs, nor will any other
  planned experiment yield as many $\theta_\e$ measurements for these
  objects as KMT.  Thus, the currently anticipated 10-year KMT survey
  will remain the best way to study bulge BDs for several decades to
  come.

\end{abstract}


\section{{Introduction}
\label{sec:intro}}

Isolated dark\footnote{Nothing in the universe is truly ``dark''.
All objects with a surface reflect ambient light, and even black holes emit
\citet{hawking75} radiation.  By ``dark'', we mean specifically: ``undetectable
with current or planned instruments''.} objects can only be studied with
gravitational microlensing.  In principle, their masses, distances, and
transverse velocities can be determined by measuring seven parameters,
which can be summarized as five quantities (two of which are vectors).
These are the Einstein timescale ($t_\e$), the angular Einstein radius
($\theta_\e$), the microlens parallax vector ($\bpi_\e$), and the source
parallax and proper motion, ($\pi_S$ and $\bmu_S$).  Here,
\begin{equation}
t_\e = {\theta_\e\over\mu_\rel}; \qquad
\theta_\e = \sqrt{\kappa M\pi_\rel}; \qquad
\kappa\equiv {4\,G\over c^2\,\au} \simeq 8.14\,{\mas\over M_\odot},
\label{eqn:tedef}
\end{equation}
and
\begin{equation}
\bpi_\e = \pi_\e{\bmu_\rel\over\mu_\rel};\qquad \pi_\e\equiv{\pi_\rel\over\theta_\e},
\label{eqn:piedef}
\end{equation}
where $M$ is the lens mass, ($\pi_\rel,\bmu_\rel$) are the
lens-source relative (parallax, proper-motion), 
and $\mu_\rel \equiv |\bmu_\rel|$.
Then, the lens mass, distance $D_L$, and transverse velocity $\bv_\perp$ are
given by
\begin{equation}
  M = {\theta_\e\over\kappa\pi_\e}; \qquad
  D_L = {\au\over \pi_\rel + \pi_S}; \qquad
  \bv_\perp = D_L(\bmu_S + \bmu_{\rel,\hel}),
\label{eqn:mdeval}
\end{equation}
where
\begin{equation}
  \pi_\rel = \theta_\e\pi_\e; \qquad
  \bmu_{\rel,\hel} = \bmu_\rel
  + {\pi_\rel\over\au}\bv_{\oplus,\perp}; \qquad
  \bmu_\rel = {\bpi_\e\over \pi_\e}\,{\theta_\e\over t_\e}
\label{eqn:mdeval2}
\end{equation}
and where $\bv_{\oplus,\perp}$ is Earth's orbital velocity at the peak
of the event projected on the sky.

After 30 years of dedicated surveys that have cataloged more than 30,000
microlensing events, such complete characterizations have been carried out
for exactly one isolated dark object: OGLE-2011-BLG-0462/MOA-2011-BLG-191,
which is a black hole (BH, \citealt{ob110462a,ob110462b}) with
mass $M=7.88\pm 0.82\,M_\odot$, distance\footnote{In their abstract,
\citet{ob110462c} quote a shorter distance based on an incorrect estimate of
$\pi_S$ that was adopted from \citet{ob110462a}, but they correct this in the
penultimate paragraph of the main body of their paper.}
$D_L= 1.62\pm 0.15\,\kpc$, and $v_\perp=43.4\pm 3.8\,\kms$,
with a direction (Galactic North through East) of $\phi=-17^\circ$
\citep{ob110462c}.

If $\bmu_S$ cannot be measured or if the direction of $\bmu_\rel$ is
ambiguous (and if $\pi_S$ can be measured or adequately
estimated), then one loses the measurement of $\bv_\perp$ but still retains
those of $M$ and $D_L$.  There are three other such mass
measurements of isolated dark objects, all brown dwarfs (BDs).
These are OGLE-2007-BLG-224, with $M=58\pm 4\,M_{\rm jup}$,
$D_L=0.52\pm 0.04\,\kpc$ \citep{ob07224} and OGLE-2015-BLG-1268,
with $M=47\pm 7\,M_{\rm jup}$ and $D_L=5.9\pm 1.0\,\kpc$ \citep{ob150763}, and
OGLE-2017-BLG-0896 \citep{ob170896}, with $M=19\pm 2\,M_{\rm Jup}$, and
$D_L= 4.0\pm 0.2\,\kpc$. % and ambiguous $v_\perp$.

For OGLE-2007-BLG-224, the lens is so much closer to the Sun than the
source and has such high $\mu_{\rel,\hel}=43\,\masyr$, that
$\bmu_S$ plays very little role in assessing the kinematics of the lens.
Moreover, its source star is in the {\it Gaia} \citep{gaia16,gaia18}
DR3 catalog.  Although it does not have a proper-motion measurement in DR3,
its proper motion
could be measured in future {\it Gaia} data releases.  
However, the other two events both have ambiguous directions for $\bmu_\rel$,
so the transverse velocities will remain unknown even if $\bmu_S$ is later
measured.

These four mass-distance determinations of isolated dark objects nicely
illustrate the three methods that have been used to date to measure $\bpi_\e$,
as well as two of the three methods that have been used to measure $\theta_\e$.
For OGLE-2011-BLG-0462, $\bpi_\e$ was measured by annual parallax
\citep{gould92}, while $\theta_\e$ was measured using astrometric microlensing
\citep{walker95,hnp95,my95}.  For OGLE-2007-BLG-224, $\bpi_\e$ was measured by
terrestrial parallax \citep{holz96,gould97}, while $\theta_\e$ was measured
from finite-source effects as the lens transited the source
\citep{gould94a,witt94,nemiroff94}.  For OGLE-2015-BLG-1268 and
OGLE-2017-BLG-0896, $\bpi_\e$ was
measured by satellite parallax \citep{refsdal66,gould94b},
while $\theta_\e$ was again measured from finite-source effects.
The other method that has been used to measure $\theta_\e$ (but not yet applied
to isolated dark objects) is interferometric resolution of the microlensed
images \citep{delplancke01,kojima1,cassan21}.

For completeness, we note that there are two other events with mass
measurements whose BD nature could be confirmed (or contradicted)
by future adaptive-optics (AO) observations 
on extremely large telescopes (ELTs).  One of these is OGLE-2015-BLG-1482
\citep{ob151482}, which has two solutions, one with a BD lens
($M=57\,M_{\rm jup}$, $D_L=7.5\,\kpc$) and the other with a stellar lens
($M=100\,M_{\rm jup}$, $D_L=7.2\,\kpc$).  If the latter is correct, then
$\mu_\rel\simeq 9\,\masyr$, so that at first AO light on ELTs, plausibly
2030, the lens and the clump-giant source will be separated by about
135 mas, which would permit the lens to be detected for the stellar-mass
solution.  Hence, if the lens is not detected, then the BD solution will
be correct.  The other is OGLE-2016-BLG-1045 \citep{ob161045}, which has a
unique solution at the star-BD boundary,
($M=0.08\pm 0.01\,M_\odot$, $D_L=5.02\pm 0.14\,\kpc$).
If the lens is a star (and so is luminous), then it can be detected
in AO observations after it is sufficiently separated from the source.
However, because the source is a giant, the proper motion is only
$\mu_\rel\sim 7\,\masyr$, and a stellar lens must be detectable down to
very faint magnitudes in order to confirm a possible BD, this may not be
feasible immediately after first AO light on ELTs, depending on the
performance of these instruments.

As we will discuss in Section~\ref{sec:discuss}, there are good long-term
prospects for obtaining complete solutions for isolated dark objects
in both regimes, i.e., both dark massive remnants and dark substellar objects.
However, while the new instruments required to make progress in the high-mass
regime are already coming on line, those required for the low-mass regime
are several years  to several decades away.  Hence, for the present,
techniques that yield only partial information are still needed to probe
substellar-object populations.

Two such methods have been developed to date: analysis of the $t_\e$
distribution of detected microlensing events \citep{sumi11,mroz17}, and
analysis of the $\theta_\e$ (and $\mu_\rel=\theta_\e/t_\e$)
distributions of the subset of single-lens
events that have such measurements, so-called finite-source point-lens (FSPL)
events \citep{kb192073,gould22}.  Each approach has its advantages and
disadvantages.  Before comparing these, it is important to note that
both approaches must rely on Galactic models to interpret their results because
the masses and distances of individual objects cannot be determined from either
$t_\e$ or $(\theta_\e,\mu_\rel)$ measurements alone.

The great advantage of the $t_\e$ approach is that very large samples of events
are available, even accounting for the fact that only fields
subjected to high-cadence observations can contribute substantially to
the detection of low-mass objects.  In particular, \citet{mroz17} leveraged
this advantage to make the first detection of six members of a
population of
short, $t_\e\sim 0.5\,$day, events, which was separated by a gap from the
main distribution of events, and which they suggested might be due to a
very numerous class of free-floating planets (FFPs).  However, while this new
population is clearly distinct in $t_\e$ space, it is difficult to constrain its
properties based on $t_\e$ measurements alone.  In particular, there is
no reason, a priori, to assume that its density distribution follows that
of the luminous stars that define the Galactic model. The events just on the
larger-duration
side of the $t_\e$ gap are almost certainly dominated by the lowest-mass
part of the stellar-BD population.  Because the luminous-star component of
this distribution can be studied by other techniques, models of the luminous
component can provide powerful constraints that facilitate disentangling the BD
signal within the $t_\e$ distribution, which is necessarily a convolution
of BD and stellar components.  Nevertheless, because the BD component may
differ substantially in the Galactic bulge and/or the distant disk relative to
the local one that can be directly studied, it may be difficult to disentangle
the different populations, given that there is little information on the mass
and distance of each lens and that these are convolved with the kinematics.
See Equation~(\ref{eqn:tedef}).

The $\theta_\e$ approach, by contrast, has several orders of magnitude fewer
events.  For example, \citet{gould22} recovered just 30 giant-source
FSPL events from 4 years of Korean Microlensing Telescope Network
(KMTNet, \citealt{kmtnet}) data compared to an underlying sample of about
12,000 events.  Nevertheless, this approach has several major advantages,
particularly in the study of low-mass objects.  The most important is that
for a source of fixed angular radius, $\theta_*$, the rate of FSPL events
scales as $\theta_*$ (i.e., is independent of lens mass, \citealt{gould13}),
whereas the rate of microlensing events in general
scales $\propto \theta_\e \sim \sqrt{M}$.  Thus,
among the 30 FSPL events found by \citet{gould22}, four were from the
same FFP population from which \citet{mroz17} identified six members from
a much larger sample.  As in the \citet{mroz17} $t_\e$ sample, these
were separated by a gap from the main body of detections, which spans a factor
of three in $\theta_\e$ and which \citet{kb192073} dubbed
the ``Einstein Desert''.  See Figure~4 from \citet{gould22}.

A second major advantage, which follows from the same $\propto\theta_*$
scaling, but is perhaps less obvious, is that FSPL events are heavily
weighted toward bulge lenses (at least for those above the Einstein Desert).
This is because, just as $\theta_\e\propto\sqrt{M}$, it is equally the
case that $\theta_\e\propto\sqrt{\pi_\rel}$.  This effect can be seen in
Figure~9 of \citet{gould22}.  Hence, in an FSPL sample (above the
Einstein Desert), one is primarily studying bulge lenses, which renders the
interpretation cleaner.

The combination of these two effects implies that the six events found
by \citet{gould22} in the range $25<\theta_\e/\muas<50$ are likely
to be overwhelmingly bulge BDs, with possibly some contamination
by very low-mass stars.  That is, scaling to a characteristic
bulge-lens/bulge-source relative parallax\footnote{Note that for the
parent population of microlensing events, the characteristic separation
is closer to $D_{LS}\sim 1\,\kpc$, corresponding to
$\pi_\rel\sim 16\,\muas$
because larger separations are favored by the $\theta_\e\propto \sqrt{\pi_\rel}$
cross section.  However, FSPL events do not have this cross-section factor.}
$\pi_\rel\sim 10\,\muas$ (corresponding to a distance along the line of
sight $D_{LS}\equiv D_S-D_L\sim 650\,\pc$),
\begin{equation}
  M = {\theta_\e^2\over\kappa \pi_\rel} =
32\,M_{\rm jup}  \biggl({\theta_\e \over 50\,\muas}\biggr)^2
  \biggl({\pi_\rel \over 10\,\muas}\biggr)^{-1}.
  \label{eqn:scale}
\end{equation}

This brings us to the next advantage, which is that it is relatively
straightforward to vet an FSPL sample of BD candidates for stellar
``contamination''.  That is, in contrast to the underlying sample
of events (with only $t_\e$ measurements), $\mu_\rel$ is known for each
FSPL event.  Hence, it is possible to know in advance both how long
one must wait for AO observations that could potentially see the lens
(assuming that it is luminous) and what is the annulus around the source
that must be investigated in the resulting image.  Such AO followup
observations would also reveal whether the BD was truly isolated.
In fact, \citet{ob07224b} carried out this test for OGLE-2007-BLG-224 and
ruled out the possibility of a main sequence lens.

Moreover, because there are relatively few BD/stellar FSPL objects in
the \citet{gould22} sample, which likely includes roughly 8 BDs, one
could afford to probe a substantially larger fraction of this sample
than just the events that are most likely to be BDs.  The lenses
that were thereby revealed to be luminous would in themselves be
useful because they could confirm (or contradict) the predictions of
the Galactic model.

That said, this third advantage is somewhat compromised for the giant-source
FSPL sample of \citet{gould22} because even the 10--15 year interval
between the events and ELT AO first light may not be enough for the
source and lens to separate sufficiently to probe to the hydrogen burning
limit.  Given the source-magnitude distribution shown in Figure~1 of
\citet{gould22}, contrast ratios of $\ga 10^4$ in the $K$ band would be
required.  While we cannot be too precise about instruments that have
not yet been built, scaling from AO on Keck, this will probably require
separations of $\ga 100\,\mas$ even on the 39m European ELT.  From Figure~5 of
\citet{gould22}, one sees that this will not occur until well after 2030
for many of the BD candidates.

Here, we present KMT-2022-BLG-2397, the second\footnote{As this paper was
being completed, \citet{moaffp} announced the first such object,
MOA-9yr-1944, with $\theta_\e=46\pm 10\,\muas$.}
FSPL bulge-BD candidate
with a dwarf-star source.  Because this source is about 4 mag fainter
than the clump, the required contrast ratio will be about 250 rather
than $\ga 10^4$.  Based on its measured proper motion,
$\mu_\rel=6.7\pm 1.0\,\masyr$, the source and lens will be separated
by about 50 mas in 2030 and 85 mas in 2035.  The first will be sufficient
to probe for typical luminous companions, while the second will be good
enough to probe to the hydrogen-burning limit.

After presenting the analysis of this event, we discuss the prospects
for identifying a larger sample of bulge-BD candidates with dwarf-star
sources in context of the ongoing development of other possible
competing methods to probe bulge BDs. Relatedly, after the initial
draft of this paper was essentially complete, \citet{moaffp} and
\citet{moaffp2} posted two papers on an FSPL search carried out
based on 9 years of MOA data and combining measurements of the $t_\e$ and
$\theta_\e$ distributions. This search is highly relevant but does not
materially impact our investigations. Hence, we discuss this work in
detail in Section~\ref{sec:moaffp}, but do not otherwise modify the main body of
the paper (except for adding footnotes 4 and 6).

\section{{Observations}
\label{sec:obs}}

KMT-2022-BLG-2397 lies at (RA, Decl)$_{J2000}=$(18:02:19.73,$-26$:53:28.10),
corresponding to $(l,b)=(+3.64,-2.15)$.  It was discovered 
using the KMT EventFinder \citep{eventfinder} system, which is applied
post-season to the data taken with KMTNet's three identical
telescopes in Australia (KMTA), Chile (KMTC), and South Africa (KMTS).
These telescopes have 1.6m mirrors and are equipped with $4\,{\rm deg}^2$
cameras.  The observations are primarily in the $I$ band, but after every
tenth such exposure, there is one in the $V$ band for the purpose of
measuring the source color.  The event lies in the overlap of KMT fields
BLG03 and BLG35.  For KMTC, these are observed at nominal cadences of
$\Gamma=2\,{\rm hr}^{-1}$ and
$\Gamma=0.4\,{\rm hr}^{-1}$, while for KMTS and KMTA,
$\Gamma=3\,{\rm hr}^{-1}$ and
$\Gamma=0.3\,{\rm hr}^{-1}$ for the respective fields.
The data were initially reduced using the KMT pySIS \citep{albrow09} pipeline,
which is a specific realization of difference image analysis
(DIA, \citealt{tomaney96,alard98}) tailored to the KMT data.  They were
re-reduced for publication using a tender loving care (TLC) version of pySIS
(Yang et al. 2023, in prep).

Note that KMT-2022-BLG-2397 was not discovered by the in-season
KMT AlertFinder \citep{alertfinder} system, which is operated once
per weekday during most of the observing season and which is responsible
for a substantial majority of all KMT event detections. As the purpose of this
system is to identify events that are suitable for follow-up observations, it is
tuned to rising events.  The KMT-2022-BLG-2397 microlensing event
was undetectable in the data
until about one hour before peak.  Hence, by the time it would normally have
been subjected to AlertFinder analysis 6 hours later, it was already falling.
Moreover, this was on a Sunday, so there was no actual analysis until
Monday, when the event was essentially at baseline.

The KMTA data from BLG35 are of very poor quality and so are not included
in the analysis.  This exclusion has almost no effect for three reasons.
First, the cadence for BLG03 is 10 times higher.  Second, the KMTA data
fall on an unconstraining part of the light curve.  Third, conditions
were poor on the one night when KMTA data would be relevant at all.

KMT-2022-BLG-2397 was recognized as a potentially interesting event during the
initial review of the roughly 500 new EventFinder events from 2022.

\section{{Light Curve Analysis}
  \label{sec:anal}}

\begin{figure}
\plotone{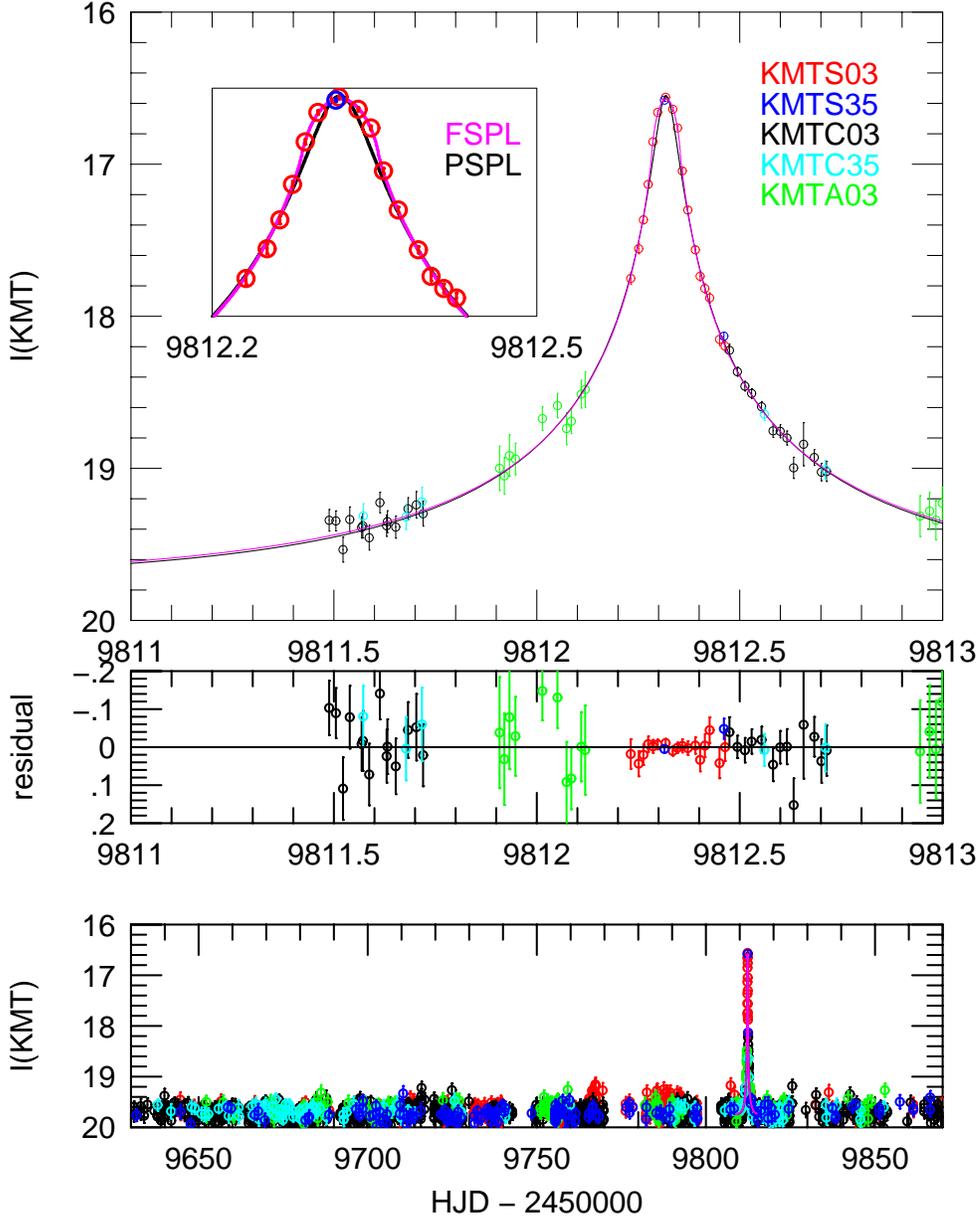}
\caption{Data (colored points) and single-lens single-source
  models for KMT-2022-BLG-2397 that do include (FSPL) or do not
  include (PSPL) finite-source effects.  The bottom panel shows the whole
  2022 season, while the top panel shows the 2 days around the peak.
  The middle panel shows the residuals to the FSPL fit.  The inset
  is a zoom of the peak region, in which it is clear that the
  PSPL model (green) fails to match the data.  Formally, $\Delta\chi^2=128$.
}
\label{fig:lc}
\end{figure}

Figure~\ref{fig:lc} shows a very short, classic FSPL event, in which
the rising light curve first steepens as the lens starts to transit the
source, and then flattens over peak, followed by a symmetric decline.
The residuals show no systematic deviations over the peak.  The peak
is captured entirely by KMTS data, with 19 points from BLG03 and 2 from BLG35,
over a total of 5.6 hours.  The inset shows a point-source point-lens (PSPL)
fit in black for comparison.  It clearly cannot match the sharp rise and
fall on the wing, nor the flattening over the peak, of the data (and the
FSPL model).  Formally it is rejected by $\Delta\chi^2=128$.  In addition,
the source magnitude according to the PSPL model
(when converted to the calibrated OGLE-III system,
see Section~\ref{sec:cmd}) is substantially brighter than the baseline
object from the OGLE-III catalog.  This is further evidence against the PSPL
model, but a full explanation would involve additional complications,
and as it is not necessary for the rejection of the PSPL model,
we do not pursue it.

Table~\ref{tab:parms}
shows the five fit parameters of the model $(t_0,u_0,t_\e,\rho,I_S)$.
Here $t_0$ is the time of the peak, $u_0$ is the impact parameter normalized
to $\theta_\e$, $\rho=\theta_*/\theta_\e$, and $I_S$ is the source magnitude
in the KMTC03 system.  In addition, we show the four ``invariant''
parameters, $t_\eff\equiv u_0 t_\e$, $t_*\equiv \rho t_\e$, $z_0\equiv u_0/\rho$,
and $f_S t_\e$, where $f_S\equiv 10^{-0.4(I_S-18)}$.

In fact, in order to emphasize several key points, we have put the
nonstandard parametrization $(t_0,t_\e,t_\eff,f_S t_\e)$ in the top
rows of Table~\ref{tab:parms}.  The first point is that the unit of
all five of these parameters is time.  Second, the last three of these
parameters each have errors of $\la 2.5\%$ compared to the $\sim 9\%$
errors of the corresponding standard parameters $(u_0,\rho,f_S)$ from
which they are derived.  Thus, these larger errors are rooted in their
correlation with $t_\e$, which is 8\%.  Finally, while the error in
$t_0$ appears to be impressively small (just 30 seconds), in fact this
does not enable any precision physical measurements.  For example,
during this seemingly short interval, the lens sweeps across the
observer plane by a distance
$(\au/\pi_\rel)\mu_\rel \sigma(t_0) \sim  0.15\,R_\oplus(\pi_\rel/\mas)^{-1}$.
Hence, for the great majority of lenses, which have $\pi_\rel\la 0.1\,\mas$,
there could not be even a $1\,\sigma$ terrestrial parallax measurement
(even assuming that the event had been observed over peak at another Earth
location).

We also note that while the error in the impact parameter $u_0$ is 9\%, 
its value relative to the source size, i.e., $z_0\equiv u_0/\rho$ is measured
to 2.4\%.
%This fact will also be important in the context of the discussion in Section~\ref{sec:discuss}.

\section{{Source Properties}
\label{sec:cmd}}

Because $\rho$ was measured, we can, in principle,
use standard techniques \citep{ob03262} to determine the
angular source radius, $\theta_*$ and so infer
$\theta_\e$ and $\mu_\rel$:
\begin{equation}
\theta_\e = {\theta_*\over \rho}; \qquad
\mu_\rel = {\theta_\e\over t_\e}.
\label{eqn:mu-thetae}
\end{equation}
In this approach, one measures the offset $\Delta[(V-I),I]$ of the
source star relative to the clump, add in the known dereddened color
and magnitude of the clump $[(V-I),I]_{\rm cl,0}=(1.06,14.34)$
\citep{bensby13,nataf13} to obtain the dereddened position of
the source $[(V-I),I]_{\rm s,0}=[(V-I),I]_{\rm cl,0}+\Delta[(V-I),I]_{S,0}$.
Then one applies a color/surface-brightness relation to determine $\theta_*$.
In our case, we use the $V/K$ relations of \citet{kervella04} by first
applying the $VIK$ color-color relations of \citet{bb88}
to transform from $V/I$ to $V/K$.

However, the practical implementation of this approach poses greater
challenges than is usually the case.  The first challenge is that there
is only one well-magnified $V$-band measurement to be used to measure
$(V-I)_S$.  This is partly a consequence of the fact that the event is
extremely short ($t_\e\sim 1.3\,$day), that the source is faint $(I_S\sim 21)$,
that it is substantially magnified for only a few hours, and that only 9\%
of the observations are in the $V$-band.  However, in this case, these
problems were exacerbated by what is essentially a bug in the
observation-sequence script.  The script alternates between a series
(roughly one per night) of 176 observations that contain no BLG02 or BLG03
$V$-band observations and another series that contains one $V$-band
observation per five $I$-band observations.  In the great majority
of cases, this makes essentially no difference because events in these
fields also lie in BLG42 or BLG43, for which the pattern is reversed.
And the great majority of the events that do not have such overlap remain
near their peak magnification for many days.  However, neither of these
``failsafes'' applied to KMT-2022-BLG-2397.  Fortunately, one of the
two BLG35 $I$-band observations was complemented by a $V$-band observation,
and it happened to be right at the peak of the event.  See Figure~\ref{fig:lc}.
This permits a measurement of the $(V-I)$ color, but without a second
observation as a check.  While there is an additional $V$-band observation
from KMTC03 on the falling wing, the source was by that time too faint to
permit a useful measurement.

\begin{figure}
\plotone{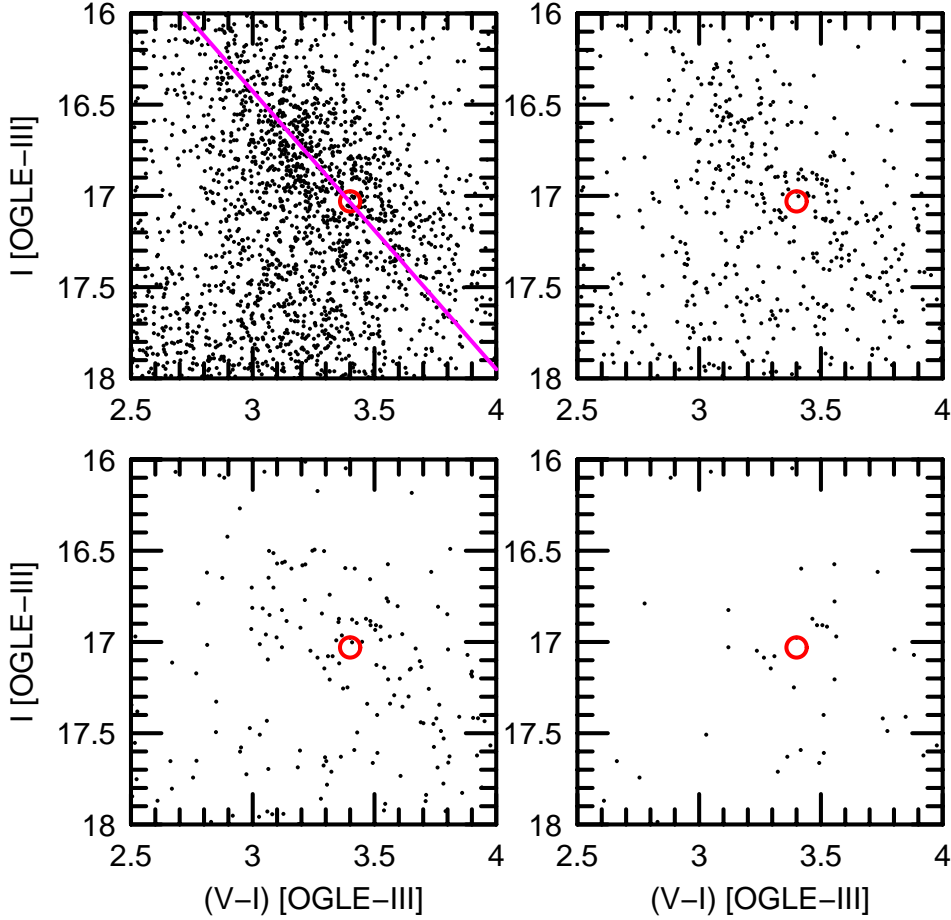}
\caption{Four views of the CMD in the region near the red clump for
  KMT-2022-BLG-2397, including stars within radii of
  $200^{\prime\prime}$ (upper left), $100^{\prime\prime}$ (upper right),
  $60^{\prime\prime}$ (lower left), and $30^{\prime\prime}$ (lower right) from
  the event.  The red circle represents the adopted centroid
  of the red clump at the location of the source star.  The magenta line
  is the reddening direction, as derived from the extension of the clump
  in the $200^{\prime\prime}$ panel.
}
\label{fig:cmdall}
\end{figure}

% 2.72 16 to 4 17.95 => d(V-I)/dI = (4-2.72)/(17.95-16) = 1.28/1.95 = 0.656
% d E(V-I)/d A_I = 0.656  R_VI = 1/0.656 = 1.52
% d\ln theta_*/d A_I = (d\ln theta_*/d I)(d A_I)+(d\ln theta_*/d(V-I))(d(V-I)) 
% = (d\ln theta_*/d I)(d A_I)+(d\ln theta_*/d(V-I))(d A_I)/R_VI
% = [(d\ln theta_*/d I) + (d\ln theta_*/d(V-I))/R_VI](d A_I)
% dln theta_*/d I = dln theta_*/[-2.5*dlog(f)] = dln theta_*/[-2.5/ln10 dln(f)]
% -ln10/5
% dln theta_*/d(V-I) = 1.4 kervella
% dln theta_*/d A_I = [-ln10/5 + 1.4/1.52] = 0.46
A second major issue is that there is an unusually large amount of
differential reddening in the neighborhood of the source.
Figure~\ref{fig:cmdall} shows the clump region of the OGLE-III \citep{oiiicat}
color-magnitude diagram (CMD), centered on the event and with radii of
$200^{\prime\prime}$, $100^{\prime\prime}$, $60^{\prime\prime}$, and $30^{\prime\prime}$.
The red circle indicates our ultimately-adopted position of the clump centroid
(see below), while the magenta line in the upper-left panel gives our
estimate of the reddening track.  It has a slope of
$R_{VI}\equiv \Delta A_I/\Delta E(V-I)=1.52$.

Within the $200^{\prime\prime}$ circle, the clump is quite extended and its
centroid is substantially brighter and bluer than our ultimately-adopted
position.  For the $100^{\prime\prime}$ circle, the clump remains extended,
but its centroid is closer to our adopted position.  For the $60^{\prime\prime}$
circle, the clump centroid is close to our adopted position, although
it is already very thinly populated.  The $30^{\prime\prime}$ circle
confirms that the clump centroid is well localized near our adopted
position, although one would not try to measure the clump position based on
this panel alone.

\begin{figure}
\plotone{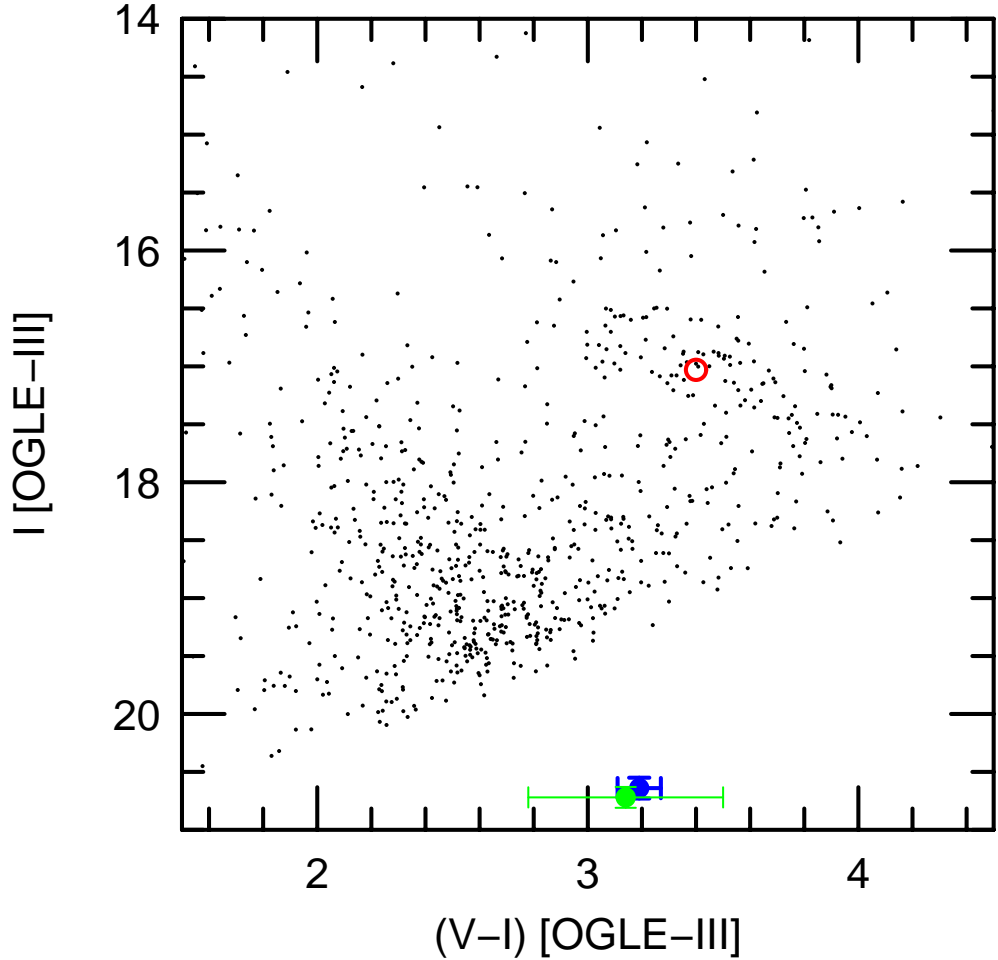}
\caption{Full CMD within $60^{\prime\prime}$ of KMT-2022-BLG-2397.
  The red circle is the clump centroid (same as in Figure~\ref{fig:cmdall}).
  The blue and green points are two independent measurements of the
  source color and magnitude, from KMTS35 and KMTC01, respectively.
  The latter provides a truly independent measurement of $I_S$ but only
  qualitative confirmation of $(V-I)_S$.
}
\label{fig:cmd}
\end{figure}

Figure~\ref{fig:cmd} shows the full CMD within the $60^{\prime\prime}$ circle.
The blue and green points represent the position of the source as
determined from the KMTS35 and KMTC03 fields respectively.  The latter
only qualitatively confirms the source color, but it does give
an independent measurement of the source magnitude.

We determined the source CMD parameters as follows.  First, we reduced these
two data sets using pyDIA \citep{pydia}, which yields light-curve and
field-star photometry on the same system.  Next, we evaluated $V_{S,\rm KMT}$
and $I_{S,\rm KMT}$ by regression on the best-fit model
from Section~\ref{sec:anal}.  Note that simple regression of the two light
curves on each other should not be used
to determine $(V-I)_S$ because the two bands are affected by different
limb darkening when the lens is transiting the source, which is true
of the one point that completely dominates the signal.  We specify that
we adopted linear limb-darkening coefficients $\Gamma_I=0.440$ and
$\Gamma_V=0.621$, corresponding to a $T=5500$K star.  However, we also note that
the difference relative to the regression method is small compared to the
statistical errors.  We then determine the transformation from each of
the two KMT instrumental systems to the OGLE-III calibrated system
by matching their respective field stars.  We find, in the OGLE-III calibrated
system, $[(V-I),I]_{\rm cl}=(3.40,17.03)\pm(0.03,0.04)$ (where we have not
yet included the effects of differential reddening), 
$[(V-I),I]_{S,\rm KMTS35}^{\rm calib}=(3.19,20.64)\pm(0.08,0.09)$, and
$[(V-I),I]_{S,\rm KMTC03}^{\rm calib}=(3.14,20.72)\pm(0.36,0.09)$.  We adopt
$[(V-I),I]_S=(3.19,20.68)\pm(0.08,0.09)$, and so
$[(V-I),I]_{S,0}=(0.85,17.99)\pm(0.09,0.10)$.

Following the above-mentioned procedures of \citet{ob03262}, 
this yields, $\theta_* = 0.92\pm 0.13\,\muas$, where we have added
5\% in quadrature to account for systematic errors in the method.

We estimate a possible additional uncertainty due to differential reddening
as follows.  If there is more (or less) extinction than we have
estimated, $\Delta A_I$, then the inferred dereddened CMD position of the source
will be brighter and bluer (or fainter and redder) than we have estimated.
The combined effect is that our estimate of $\theta_*$ would then be displaced
by
\begin{equation}
  {d\ln\theta_*\over d A_I} = {\ln 10\over 5}-{d\ln\theta_*/d(V-I)_0\over R_{VI}}
    \rightarrow -0.46,
  \label{eqn:AIeffect}
\end{equation}
where we have evaluated $d\ln\theta_*/d(V-I)_0=1.4$ using the same procedures as
above.  Based on Figure~\ref{fig:cmdall}, we estimate $\sigma(A_I) = 0.1$
and therefore a contribution to $\sigma(\ln\theta_*)$ of 4.6\%.  Adding this
in quadrature, we finally adopt $\theta_* = 0.92\pm 0.14\,\muas$, and hence
\begin{equation}
  \theta_\e = 24.8 \pm 3.6\,\muas; \qquad
  \mu_\rel = 6.69 \pm 0.96\,\masyr .
  \label{eqn:thetaemu}
\end{equation}

Next, we compare the source star to the baseline object as given by the OGLE-III
catalog, in terms of both flux and astrometric position.
First, the magnitude of the
source on the OGLE-III system is $I_S=20.68\pm 0.09$, while the baseline object
from the OGLE-III catalog has $I_{\rm base} = 20.58$.  That is, there is no
evidence for blended light.  In contrast to the situation for the source,
the error on the baseline flux is driven primarily by surface brightness
fluctuation due to undetected faint stars.  Hence, blended light of
$f_B\la 0.3\,f_S$ cannot be ruled out.  If the putative blend were in the
bulge, then this limit would still permit all main-sequence stars with masses
$M_B\la 0.9\,M_\odot$.  Therefore, this limit is only mildly constraining.

We transform the source position (derived from centroiding the
difference images of the magnified source) to the OGLE-III system
for each of the KMTC03 and KMTS35 reductions.  These differ by
$\Delta\btheta(E,N) = (24,27)\,\mas$, which leads to a rough estimate of
the combined error of the pyDIA measurements and transformation of
$\sigma\simeq 18\,\mas$ for each measurement.
Then comparing the average of these
two determinations with the position of the baseline object, we find
$\btheta_{\rm base} - \btheta_{S} = (76,50)\,\mas$.  This difference is much larger
than the 13 mas standard error of the mean of the two source-position
measurements.  In principle, the difference could be due
to astrometric errors in the OGLE-III measurement of this faint star,
which is also affected by surface brightness fluctuations.  However,
it is also possible that the baseline object has moved by $\sim 90\,\mas$
relative to the bulge frame during the 16 years between the epoch of the
OGLE-III catalog and the time of the event.  In brief, all available
information is consistent with the baseline object being dominated by
light from the source.

\section{{Nature of the Lens}
  \label{sec:physical}}

In principle, the lens could lie anywhere along the line of sight,
i.e., at any lens-source relative parallax, $\pi_\rel$.  Applying the
scaling relation Equation~(\ref{eqn:scale}) to the result from
Equation~(\ref{eqn:thetaemu}) yields,
\begin{equation}
  M = {\theta_\e^2\over\kappa \pi_\rel} =
8\,M_{\rm jup}  
  \biggl({\pi_\rel \over 10\,\muas}\biggr)^{-1}.
  \label{eqn:scale2}
\end{equation}
Thus, if the lens is at the characteristic $\pi_\rel=10\,\muas$ of bulge lenses
(for FSPL events),
then it is formally a ``planet'' in the sense that it lies below the
deuterium-burning limit.  However, as the expected distribution of FSPL
bulge-bulge microlensing events is roughly uniform in $\pi_\rel$, it could
also be more massive than this limit (so, formally, a ``BD'').

On the other hand, if the relative parallax had a value more typical
of disk lenses, $\pi_\rel\ga 50\,\muas$, then the lens mass would be
$M\la 1.6\,M_{\rm jup}$, i.e., clearly planetary.

Nevertheless, the main interest of this object is how it relates to 
the \citet{gould22} statistical sample of FSPL events.  The fact that
KMT-2022-BLG-2397 is right at the upper shore of the Einstein Desert
suggests that it is part of the dense population of objects lying just
above this shore.  As discussed in Section~\ref{sec:intro} with respect
to Equation~(\ref{eqn:scale}), these are likely to be primarily (or entirely)
bulge BDs.  The fact that their distribution is suddenly cut off implies
a steeply rising mass function.  Regardless of whether the threshold for
this rise is above or below the deuterium-burning limit, its existence
points to a formation mechanism that is distinct from planets, including
FFPs.

Therefore, the main question regarding KMT-2022-BLG-2397 is not its 
exact nature, but rather whether and how the ensemble of objects like it,
i.e., low-$\theta_\e$ FSPL events with dwarf-star sources, can contribute
to our understanding of the BDs and FFPs that lie concentrated,
respectively, above and below the Einstein Desert.

\section{{Limits on Hosts}
  \label{sec:limits}}

If the BD candidate had a host that was sufficiently close, it could leave
trace features on the light curve, either a long-term ``bump'' directly
due to the host or subtle distortions to the FSPL profile.
We search for evidence of such features by a grid search over the three
additional parameters required to describe binary-lens systems,
$(s,q,\alpha)$.  Here, $q$ is mass ratio of the two components, $s$ is their
projected separation in units of their combined Einstein radius (which is
$\sqrt{q+1}$ larger than the Einstein radius associated with the FSPL event),
and $\alpha$ is the angle between the lens-source relative motion and the
binary axis.

\begin{figure}
\plotone{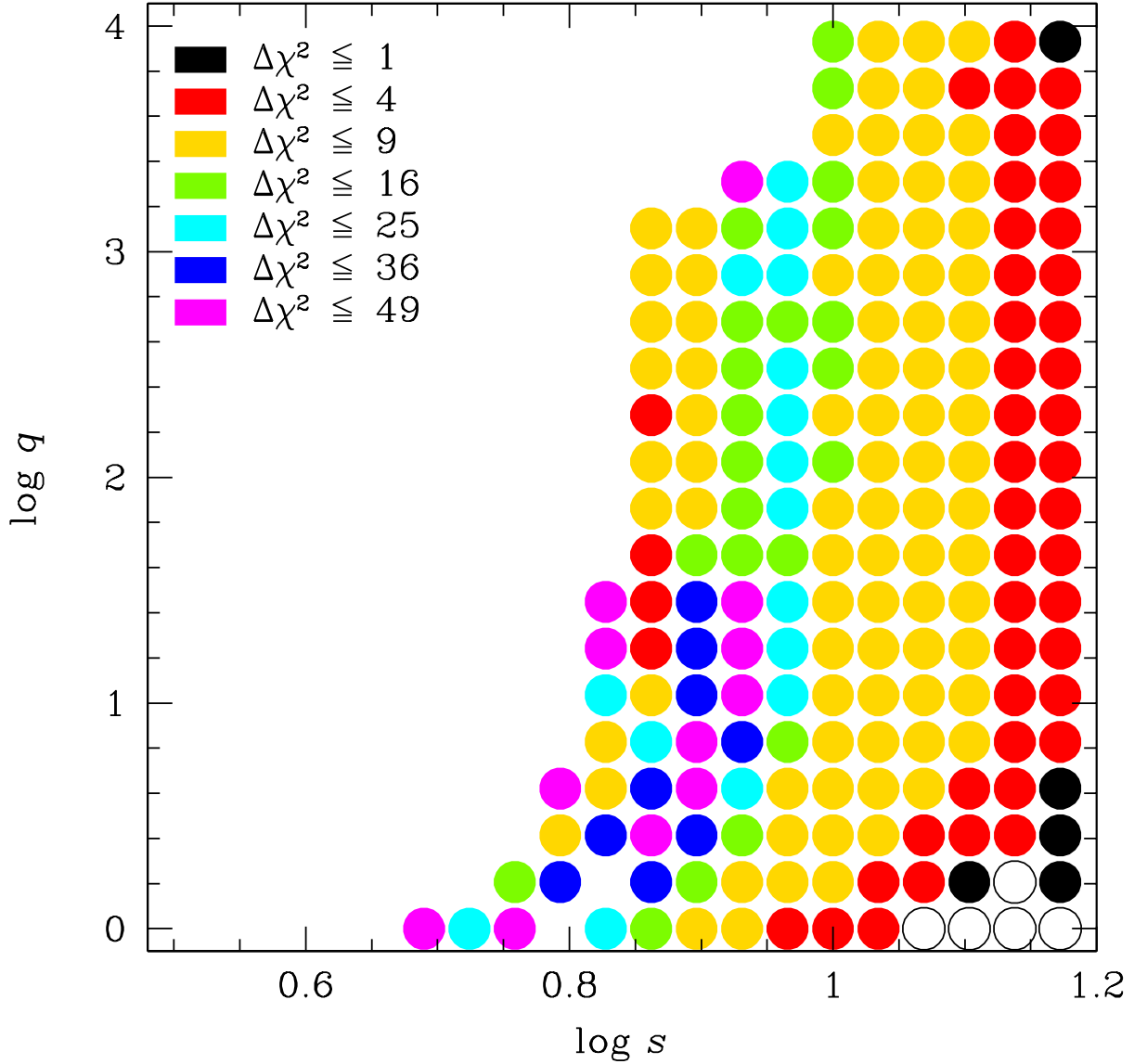}
\caption{Grid search for binary-lens solutions, with resulting
  $\Delta\chi^2 = \chi^2 ({\rm binary}) - \chi^2({\rm FSPL})$ coded by
  point type and color.  Filled symbols indicate $\Delta\chi^2>0$, while
  open symbols indicate $\Delta\chi^2<0$.  The color coding is indicated
  in the legend, while the blank regions have $\Delta\chi^2>49$.  The
  grid search is over three variables $(s,q,\alpha)$ but only the best
  $\chi^2(s,q)$ (i.e., marginalized over $\alpha$) is shown.  Values
  $\log s\leq 0.8$ ($s<6.3$) are clearly ruled out, but there is an
  ``island'' of allowed solutions just above this value.  Full exploration
  of this island yields $\Delta\chi^2\sim -2$ for 3 dof, which is not
  significant.
}
\label{fig:grid}
\end{figure}

We find that all such hosts with $s<6.3$ are excluded.  In fact, many
hosts with $s\la 10$ are excluded, but there is a small ``island'' near
$(s,q)\sim(6.6,20)$ that cannot be excluded (see Figure~\ref{fig:grid}),
and is nominally favored (after a full parameter search seeded at the
grid-point values) by $\Delta\chi^2\sim -2$ for 3 degrees of freedom.
Because this is
less than the improvement expected from pure Gaussian noise, it cannot
be regarded as evidence in favor of a host.

The projected separation of this putative host would be approximately
$\Delta\theta\simeq s\sqrt{q+1}\theta_{\e,\rm FSPL}=0.75\,\mas$.  Hence,
if this putative host were detected in future high-resolution imaging
(after the source and lens have separated on the sky), it would probably
not be possible to distinguish its position from that of the FSPL object.
Hence, it would not be possible to rule out that the detected star was
the FSPL object rather than its host.  For example, if the detected star
were a late M dwarf in the bulge, $M_{\rm star}=0.15\,M_\odot$, it could be either
the host of the FSPL object, in which case the latter would have
mass $M_{\rm FSPL}\sim 8\,M_{\rm Jup}$ and with a very typical $D_{LS}\sim 0.7\,$kpc,
or it could be the FSPL object itself, in which case it would have a very
atypical $D_{LS}\sim 30\,\pc$.

Here we merely mention these possibilities in order to alert future
observers to their existence.  Unless and until there is a detection of
stellar light that is close to the position of the FSPL object, it is
premature to speculate on its interpretation.

\section{{Discussion}
\label{sec:discuss}}

KMT-2022-BLG-2397 was discovered serendipitously, i.e., in the course
of by-eye perusal of the 2022 EventFinder sample.  In contrast to the
10 FSPL giant-source events with $\theta_\e<50\,\muas$ summarized by
\citet{gould22}, it is not part of a systematic sample and therefore
cannot be used to make statistical statements about the underlying
populations of dark isolated objects.  As conducting such systematic searches
requires vastly greater effort than finding and analyzing some interesting
events, it is appropriate to ask whether such a sample is likely to be
worth the effort.  This question can be broken down into three parts:
\begin{itemize}
\item{would such a survey likely contribute substantially to the total number
  of such small-$\theta_\e$ events? (Section~\ref{sec:relative})};
\item{would they contribute qualitatively different information relative to
  the existing giant-source sample? (Section~\ref{sec:qualitative})};
\item{is it likely that the enhanced numbers and/or improved quality could be
obtained before better experiments come on line to attack the same
underlying scientific issues? (Section~\ref{sec:context})}
\end{itemize}

Before addressing these questions, we note that of the 10 small-$\theta_\e$
giant-source FSPL events from \citet{gould22}, five were published either
before or independent of the decision by \citet{kb192073} and \citet{gould22}
to obtain a complete sample of giant-source FSPL events.  These included
two  of the four FFP candidates
(OGLE-2016-BLG-1540 and OGLE-2019-BLG-0551, \citealt{ob161540,ob190551}) and
three of the six BD candidates
(MOA-2017-BLG-147, MOA-2017-BLG-241, \citealt{mb19256}, and
OGLE-2017-BLG-0560, \citealt{ob121323}).  Thus, serendipitous detections can
play an important role in motivating systematic searches.

\subsection{{Relative Detectability of FSPL Events From Dwarf and Giant Sources}
\label{sec:relative}}

We begin by assessing the relative contributions of microlensing
events with giant sources to those whose sources are main-sequence or
subgiants (hereafter collectively referred to as “dwarfs”).
We must start with events that meet three conditions:
\begin{itemize}
\item{they are actually detected by KMT (Section~\ref{sec:snrat}),}
\item{they objectively have the property that the lens transits the source
  (independent of whether there are any data taken during this interval;
  Section~\ref{sec:relnum}), and}
\item{$\rho$ is measurable in the data (Section~\ref{sec:rho-measurable}).}
  \end{itemize}
These three conditions interact in somewhat subtle ways, so their
investigation overlaps different sections and the divisions indicated
above are only approximate. They are combined in Section~\ref{sec:7.1.sum}.

For the moment,
we simply report the result that, with respect to BD FSPL events, dwarfs
are favored over giants by a factor $\sim 2.7$, while this factor is somewhat
less for FFP FSPL events.

\subsubsection{{Is $\rho$ Measurable?}
\label{sec:rho-measurable}}

The first consideration is whether or not the data stream contains
adequate data points to measure $\rho$. For dwarfs,
the chance that the data stream will contain points that are close
enough to the peak to permit a $\rho$ measurement is substantially smaller,
simply because the duration of the peak is shorter by a factor
$t_{*,\rm d}/t_{*,\rm g}=\theta_{*,\rm d}/\theta_{*,\rm g}\sim 1/10$.
For example, 11 of KMT's 24 fields have cadence $\Gamma= 0.4\,{\rm hr}^{-1}$
and 3 have cadence $\Gamma= 0.2\,{\rm hr}^{-1}$, but for a $2t_*\sim 20\,$hr
peak, these cadences
can be quite adequate to measure $\rho$.  Indeed, five of the 30
giant-source FSPL events of \citet{gould22} came from these low-cadence fields,
despite their dramatically lower overall event rate.  See their Figure~2.
More strikingly, 15 of the 30 came from the seven fields with
$\Gamma= 1\,{\rm hr}^{-1}$.  These would also have adequate coverage for events
with dwarf-source events, provided that there were no gaps in the data
due to weather or shorter observing windows in the wings of the season.
To account for this, we estimate that one-third of these
$\Gamma= 1\,{\rm hr}^{-1}$ events would be lost due to gaps.  Dwarf-star
source would be most robustly detected in the three prime fields, which have
cadences $\Gamma=2$--4$\,{\rm hr}^{-1}$, but these fields accounted for only 10
of the 30 giant-source FSPL events.
Thus, one can expect that about two-thirds as
many dwarf-source events would have adequate coverage compared to giant sources
(relative to the numbers of actually detected events that have the property
that the lens transits the source, as discussed in the first paragraph).

\subsubsection{{Signal-to-noise Ratio}
\label{sec:snrat}}

The second consideration is that in most cases (with one important exception),
the overall signal-to-noise ratio (S/N) is lower for
a dwarf source compared to a giant source for the ``same'' event, i.e.,
same parameters $(t_0,t_\e,z_0,\theta_\e)$.  To elucidate this issue
(as well as the one aspect for which dwarf sources have a clear advantage,
see below), we follow \citet{ob190551} and analyze the signal in terms of
the mean surface brightness $S=f_S/\pi\theta_*^2$ of the source.
For purposes of illustration, we ignore limb darkening and consider the
signal from an observation when the lens and source are perfectly aligned
as representative. Hence, $A_\max=\sqrt{1 + 4/\rho^2}$, and so the excess flux
of the magnified image, $\Delta F_\max = (A_\max -1)f_S$, which can be
approximated,
\begin{equation}
  \Delta F_\max = 2\pi S \theta_\e^2 \quad (\rho\gg 1),
  \qquad
  \Delta F_\max = 2\pi S \theta_\e\theta_* \quad (\rho\ll 1).
  \label{eqn:signal}
\end{equation}
As giants are bigger than dwarfs, i.e., $\rho_{\rm g}/\rho_{\rm d}>1$, there
are three cases to consider: (1) $1>\rho_{\rm g}>\rho_{\rm d}$;.
(2) $\rho_{\rm g}>1>\rho_{\rm d}$; (3) $\rho_{\rm g}>\rho_{\rm d}>1$.
These yield ratios,
\begin{equation}
\biggl({\Delta F_{\max,\rm d}\over \Delta F_{\max,\rm g}}\biggr)_1 =
      {S_{\rm d}\over S_{\rm g}} {\theta_{*,\rm d}\over \theta_{*,\rm g}}.
\quad
\biggl({\Delta F_{\max,\rm d}\over \Delta F_{\max,\rm g}}\biggr)_2 =
      {S_{\rm d}\over S_{\rm g}}\rho_{\rm d};
\quad
  \biggl({\Delta F_{\max,\rm d}\over \Delta F_{\max,\rm g}}\biggr)_3 =
        {S_{\rm d}\over S_{\rm g}};
  \label{eqn:signalrat}
\end{equation}
To a good approximation\footnote{This is essentially the same
approximation that underlies linear color-color relations in this
regime, i.e., that the Planck factor is well-approximated by the
Boltzmann factor.}, the surface-brightness ratio in
Equation~(\ref{eqn:signalrat}) is given by
\begin{equation}
  {S_{\rm d}\over S_{\rm g}} =
  \exp\biggl[{hc\over k\lambda_I}\biggr({1\over T_{\rm g}} - {1\over T_{\rm d}}
  \biggr)\biggr] \rightarrow 2.0.
  \label{eqn:sbrat}
\end{equation}
where $\lambda_I = 810\,$nm and where we have made the evaluation at
representative temperatures $T_{\rm d}=5800\,$K and $T_{\rm d}=4700\,$K
for dwarfs and giants, respectively.
Thus, for $\rho_{\rm g}< 1$, the signal is about 5 times
larger for giants than dwarfs.  That is, the source is 10 times larger
but the surface brightness is 2 times smaller.
This is the regime of essentially
all of the FSPL events from \citet{gould22}, except for the four FFPs.
The signals only approach equality for $\rho_{\rm d}\sim 0.5$
(i.e., $\rho_{\rm g}\sim 5$).  To date, the only FSPL events near or below
this regime are the FFP candidates,
OGLE-2012-BLG-1323 ($\rho_{\rm g}=5.0$ \citealt{ob121323}),
OGLE-2016-BLG-1928 ($\rho_{\rm g}=3.4$ \citealt{ob161928}), and
OGLE-2019-BLG-0551 ($\rho_{\rm g}=4.5$ \citealt{ob190551})\footnote{Recently,
\citet{moaffp} have announced an FSPL FFP, MOA-9yr-5919, with 
$\theta_\e=0.90\pm 0.14\,\muas$.  The source is a subgiant,
$\theta_*=1.26\pm 0.48\,\muas$, so $\rho=1.4$.  However, the underlying
object can be considered to be in this regime because if the source had
been a giant ($\theta_*\sim 6\,\muas$), then $\rho\sim 6.7$.}.

Finally, giants have an additional advantage that the duration of the peak
region is 10 times longer,
so that (at fixed cadence) there are 10 times more data points, which
is a $\sqrt{10}\sim 3$ advantage in S/N.

However, when we considered the signals from excess flux $\Delta F$, we ignored
the fact that the giant signal is more degraded by photon noise compared
to the dwarf signal, simply because the baseline giant flux is greater.
This is the ``important exception'' mentioned above.  Nevertheless,
as we now show, while the importance of this effect depends strongly on the
extinction, for typical conditions it is modest.

For typical KMT seeing ($\fwhm\sim 4$ pixels) and background ($B\sim 800$ flux
units per pixel), keeping in mind the KMT photometric zero point of
$I_{\rm zero}=28$, and in the Gaussian PSF approximation, the baseline source
flux and background light contribute equally to the noise at
$I_S = I_{\rm zero} - 2.5\log(4\pi B\times\fwhm^2/\ln(256)) = 16.8$.  Given typical
extinction levels, $1\la A_I\la 3$, dwarf (including subgiant) stars are almost
always fainter than this threshold.  On the other hand, clump giants
$(I_{S,0}\sim 14.5)$ at typical extinction ($A_I\sim 2$) have about equal
photon noise from the source and the background, implying a reduction of
$\sqrt{2}$ in S/N.  Only very bright giants suffer from
substantially greater noise, but these also have a far greater signal than
the typical estimates given above.  Hence, the higher noise from giants
does not qualitatively alter the basic picture that we presented above.

In brief, for fixed conditions, the signal from the ``same'' event is
substantially greater for giant sources than dwarf sources, except for the
case $\theta_\e\la 2\theta_{*,\rm g}\sim 12\,\muas$, for which a declining
fraction of the giant is effectively magnified.  This is the regime of FFPs.

\subsubsection{{Relative Number of Events}
\label{sec:relnum}}

\begin{figure}
\plotone{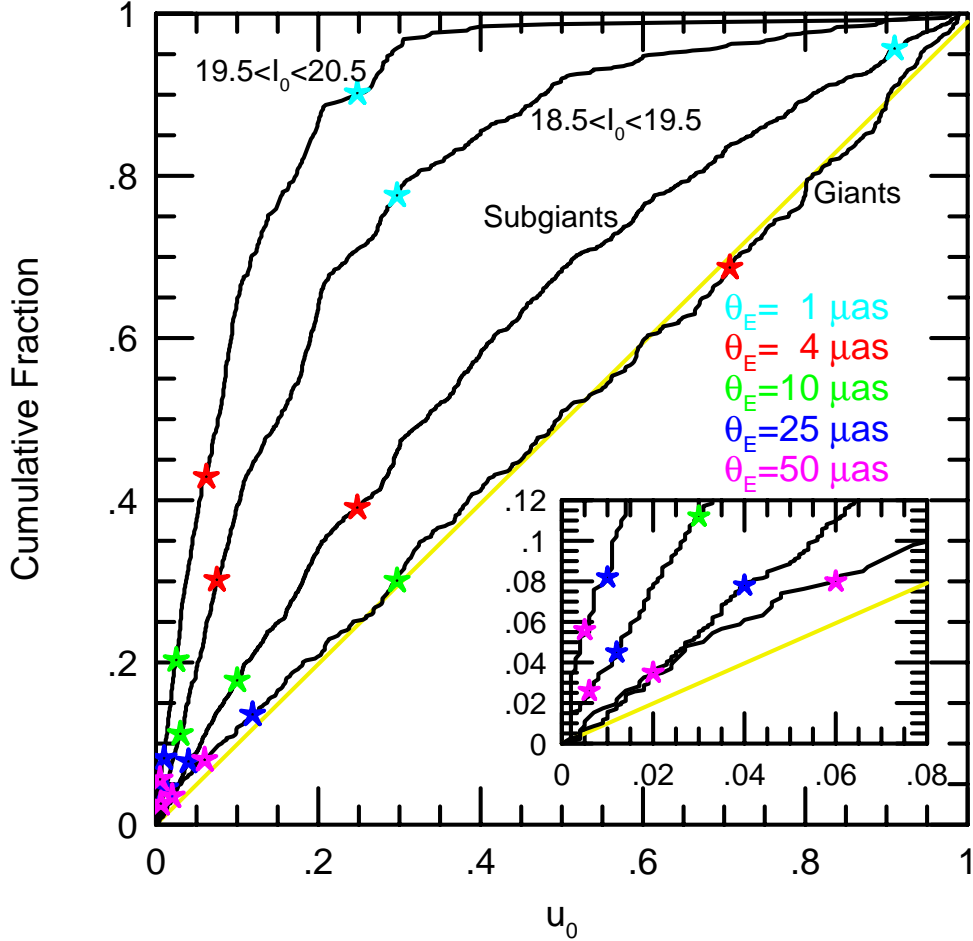}
\caption{Cumulative distributions of 2019 KMT events as a function of
  impact parameters $0<u_0 \leq 0.99$ as derived from the KMT web page
  for four sub-populations, including two groups of main-sequence stars
  ($19.5<I_0<20.5$ and $18.5<I_0<19.5$), subgiants ($16.5<I_0<18.5$),
  and giants ($13.0<I_0<16.5$).  The yellow line shows the expected
  distribution for events unaffected by selection bias, which is a
  good match to the giants.  The other populations show increasing
  evidence of magnification bias.  The colored points are at the $u_0$
  values that generate the same magnification as an FSPL event with
  $\rho=\theta_*/\theta_\e$, i.e., $u_0^2 = \sqrt{4 + \rho^2} - 2$.
  Here, we adopt $\theta_*=(0.5,0.6,2,6)\,\muas$ for the four
  sub-populations, while the $\theta_\e$ values are given in the
  legend.  For BDs ($25\la \theta_\e/\muas\la 50$), all populations
  are in the linear regime (see inset), so magnification bias plays
  essentially no role.  This remains basically the case for FFPs
  similar to those in the \citet{gould22} giant-source study
  ($4\la \theta_\e/\muas\la 10$).
  However, for substantially smaller $\theta_\e\sim 1\,\muas$,
  magnification bias plays a major role (cyan points).}
\label{fig:cumul}
\end{figure}

We close by examining the interplay between cross section, surface density,
and magnification bias as they affect the relative number of giant-source and
dwarf-source FSPL events, i.e., events that are both in the KMT sample and
have the objective property that the lens transits the source.
Figure~\ref{fig:cumul} shows cumulative distributions of $u_0$ for four
classes of events drawn from the 2019 KMT web page:
two groups of upper main-sequence stars, $19.5<I_0<20.5$ and $18.5<I_0<19.5$,
$16.5<I_0<18.5$ (``subgiants''), and $13<I_0<16.5$ (``giants'').
The parameters are derived from
the automated fits of the KMT webpage.  Events with $u_0\geq 1$ are excluded
because the automated fitter just assigns these $u_0=1$.  In addition, events
with no tabulated extinction are also excluded.  The four groups contain,
respectively, 382, 726, 1286, and 459 events for a total of 2853.
Additionally, there are 13 events with $I_0<13$ and 280 others with
$I_0>20.5$ that are excluded from this study in order to keep it simple.

The first point to note is that the giant-star sample is perfectly
consistent with being uniform, as is rigorously expected for the
underlying population of microlensing events,  That is, the maximum
difference between the giant-star curve and the yellow line is $D=0.0344$,
yielding a Kolmogorov-Smirnov (KS) statistic $D\sqrt{N}=0.74$.  The other
curves all display ``magnification bias'': they are uniformly distributed
in $u_0$ up to a point but then bend toward the right.  The respective
break points for the three classes (fainter to brighter) are roughly
$u_0\sim(0.05,0.10,0.20)$.  This is important because BD FSPL events
take place at relatively high magnification, so the relative paucity of
detected dwarf-source events at low magnification plays very little role.
% slp giant 459/1.000, subg 437/0.2 = 2185; 285/0.1 = 2850; 143/0.05 = 2860
% 6.23:6.21:4.76:1

This feature is illustrated by the magenta and blue points, which
represent the effective $u_0$ equivalents for BDs at the boundaries
of the region of interest.  The lower boundary ($25\,\muas$) is the
upper shore of the Einstein Desert, while the upper boundary ($50\,\muas$)
is an approximate upper limit for relatively secure BD candidates.  To
make these identifications, we first assign a representative
$\theta_*=(0.5,0.6,2.0,6.0)\,\muas$ to the four populations and then
equate peak magnifications, i.e., $A=\sqrt{1 + 4/\rho^2}$, and
$A=(u_0^2+2)/u_0\sqrt{u_0^2+4}$.  In other words,
$u_{0,\eff}^2 = \sqrt{4 + \rho^2}-2$.  That is, we are assuming that the
structures seen in Figure~\ref{fig:cumul} are due to peak-magnification bias.

From Figure~\ref{fig:cumul}, the BD range is entirely in the linear regime
for each of the 3 non-giant populations, while for giants, the entire
distribution is linear.  This means that the contributions of these
populations can be estimated based on the slopes in these regimes.  In other
words, the detection of BDs would be exactly the same as if these regimes
remained linear up to $u_0=1$.  We find that from faint to bright, the
linear regimes of the four populations are in ratios of 6.2:6.2:4.8:1
(i.e., the observed low-$u_0$ slopes of the normalized distributions
from Figure~\ref{fig:cumul} multiplied by the total population of each group).
Multiplying these relative source frequencies by the $\theta_*$ values
(i.e., cross sections) listed above yields ratios of relative rates
of 0.5:0.6:1.6:1. Hence, the ratio of rates of dwarf to giant
source events is $(0.5+0.6+1.6) = 2.7$.
Taking account of the cadence-induced factor of 2/3 for the effectiveness of
dwarf searches, as estimated above, the dwarfs have an overall
advantage of $(0.5+0.6+1.6)/1.5\rightarrow 2.7/1.5 = 1.8$ relative to giants.

For FFPs the situation is somewhat more complicated.  The green
and red points represent the lower shore of the Einstein Desert
$(\theta_\e=10\,\mas)$ and the smallest Einstein radius in the \citet{gould22}
sample $(\theta_\e=4\,\mas)$, respectively.
Within this range, the four distributions
are essentially in the linear regime, and hence the same argument given for
BDs still basically applies.  However, dwarf sources are potentially sensitive
to yet smaller Einstein radii, i.e., $\theta_\e<4\,\muas$, which correspond to
an FFP population that is not detectable with giants.  These are located at
positions to the right
of the red points on each curve.  Because the curves start to turn over
in these regions, sensitivity is lost relative to the approximately
linear regimes to the left of the red points.  This is particularly so
for the two main-sequence populations.  Nevertheless, substantial
sensitivity remains until approximately $\theta_*\sim 1\,\muas$ (cyan points),
beyond which the cumulative curves flatten, implying that the sensitivity
declines catastrophically.  In brief, within the $\theta_\e$-range of the
FFPs probed by giant sources, the dwarf-to-giant-source ratio will be somewhat
lower than for BDs because the curves in Figure~\ref{fig:cumul} begin to
deviate from linear.  However, in contrast to the situation for the
BDs, the dwarfs open up additional (and poorly characterized) parameter
space for the FFPs.  Hence,
we expect that the BD and FFP relative dwarf-to-giant sensitivities are
similar, while recognizing that the latter is more uncertain.

It is of some interest to compare the slope ratios derived above for the
$u_0\ll 1$ regime (6.2:6.2:4.8:1) with what would be expected based on the
relative number of sources as determined from the \citet{holtzman98}
luminosity function (HLF), based on {\it Hubble Space Telescope (HST)}
images of Baade's Window.  The HLF is effectively displayed only for
$M_I>-0.125$, corresponding to $I_0>14.375$.  Thus to make the comparison,
we first impose this restriction on the KMT giant bin, which reduces it
from 459 to 404.  For reasons that will become clear, we normalize
to the subgiants, rather than the giants.  Then, the observed ratios
are (1.31:1.30:1:0.18).
By contrast, for the HLF, we find ratios (1.94:1.45:1:0.10).
If we ignore the giants for the moment, then the following narrative
roughly accounts for the relationship of these two sets of ratios:
The KMT EventFinder and AlertFinder algorithms search the ensemble of
difference images for microlensing events at the locations of cataloged
stars.  The great majority of subgiants are in the catalog, so their
locations are searched and thus the great majority of high-magnification
events are found.  Most of the stars $18.5<I_0<19.5$ are also in the catalog,
unless they happen to be close to brighter stars, in which case their
locations are also searched.  However, some of these stars are far from
any cataloged stars and still are not included in the underlying catalogs.
Hence, the expected ratio (1.45:1) and observed ratio (1.30:1) are similar,
but there is a slight deficit for the latter due to events concurring at
unsearched locations.  Then the same argument predicts that this shortfall
will be greater for the $19.5<I_0<20.5$ because these are much less likely
to enter the catalogs even if they are isolated.  Unfortunately, this
narrative does not account for the discrepancy between expected and
observed for giants, which should enter the catalogs similarly to subgiants.
We conjecture that the narrative is essentially correct and that the
giant-subgiant comparison suffers from some effect that we have not
identified.  This could be investigated by running the EventFinder algorithm
more densely, say at $0.5^{\prime\prime}$ steps, rather than just at the positions
of cataloged stars.  This would be prohibitive
for the full $\sim 100\,{\rm deg}^2$ survey, but might be possible on small
subset.

\subsubsection{{Summary}
\label{sec:7.1.sum}}

To summarize overall, BD and FFP lenses are about 2.7 times more likely
to transit the source in main-sequence-star and
subgiant (collectively ``dwarf'') cataloged events compared to cataloged
giant-source events.  However, they are roughly two-thirds as likely to have
adequate data over peak, and are somewhat more difficult to characterize
due to lower signal.  Thus, after applying a ``characterization penalty'' to the
factor of 1.8, we find that
they are likely to contribute at perhaps 1.5 times
the rate of giants to the overall detections of substellar FSPL events.

\subsection{{Qualitatively Different Information?}
\label{sec:qualitative}}

The main potential qualitative advantage of dwarf sources over giant sources for
FSPL events is in the regime of FFPs.  As shown in Section~\ref{sec:relative},
the S/N for dwarf sources is comparable or higher on an
observation-for-observation comparison for the ``same'' event.  Formally,
in the extreme limit (case (1) from Equation~(\ref{eqn:signalrat})), this
is only a factor 2 in higher signal, with a typical further improvement
of a factor of $\sqrt{2}$ from lower noise.  And hence, this advantage is
approximately canceled by the fact that giants have $\sim 10$ times more
observations for the same cadence.  However, in the extreme regime
$\rho_{\rm g}\ga 7$ (i.e., $A_\max\la 1.04$), it may not even be possible to
recognize, let alone robustly analyze, a giant-source event because of
confusion with potential source variability.  Indeed, as of today, there
are no such FFPs that have yet been identified.

Thus, the first potentially unique feature of dwarf sources is their ability
to probe to smaller $\theta_\e$.  Indeed, in this context, it is notable
that the source of the smallest-$\theta_\e$ FSPL event to date,
OGLE-2016-BLG-1928 $(\theta_\e=0.84\pm0.06\,\muas)$, is a low-luminosity
giant, $I_{S,0}=15.8$ ($\theta_*=2.85\pm 0.20\,\muas$), i.e., 1.4 mag
below the clump.

A second potential advantage, as discussed in Section~\ref{sec:intro},
is that dwarf-source FSPL events can be subjected to AO follow-up observations
much earlier than giant-source events.  Such observations are critically
important for FFPs in order to determine whether they are truly
``free floating'' or they are in wide orbits around hosts that remain
invisible under the glare of the source as long as the source and FFP
stay closely aligned.  This issue is also relevant to BDs.  Moreover,
for BD candidates, one would like to confirm that they are actually
BDs, i.e., that their small values of $\theta_\e=\sqrt{\kappa M \pi_\rel}$
are actually due to small $M$ rather than small $\pi_\rel$.

For the case of BD candidates, one might ask how one could distinguish between
two competing interpretations of the detection of stellar light associated
with the event, i.e., that it comes from the lens itself (that is, the lens
is actually a star with small $\pi_\rel$) or a host to the lens.

This brings us to a third potential advantage of dwarf sources.  The
fact that the peak can have greater structure (caused by much smaller $\rho$)
makes it easier to detect signatures of the host during the event.
At the extreme end, such events may be dominated by this structure
rather than finite source effects, as in the cases of MOA-bin-1
\citep{moabin1} and MOA-bin-29 \citep{moabin29}. But even if host effects are
not observed, in principle, stronger limits can be set on companions (hosts) as
a function of mass ratio and separation.
Nevertheless, we showed
in Section~\ref{sec:limits} that, for the case of KMT-2023-BLG-2397, it
would probably not be possible to distinguish between two hypotheses
(lens or companion to the lens)
if there were a future detection of stellar light at the position of the event.

\subsection{{Context of Competing Approaches}
\label{sec:context}}

From the above summary, a systematic search for FSPL events with
dwarf-star sources could plausibly contribute about 1.5 times as many
measurements as the KMT giant-source search \citep{gould22}, which found
4 FFP candidates and 6 excellent BD candidates
(defined as $\theta_\e<50\,\muas$) in a 4-year search.  Hence, plausibly,
the full sample could be increased by a factor $\sim 6$ by 2026.  To the
best of our knowledge, there will be no competing approaches that yield
either more or qualitatively better information on these classes of
objects on this timescale.  Moreover, the part of this parameter space
that is of greatest current interest, i.e., FFPs, is also the part that
has the greatest unique potential for dwarf-star sources.  Therefore,
on these grounds alone, it appears worthwhile to conduct such searches.

However, on somewhat longer timescales, there are several competing
approaches that are either proposed or under development.  We 
review these as they apply to dark isolated objects, with a focus on
FFPs and BDs.

\subsubsection{{Prospects for Isolated-Object Mass-Distance Measurements}
\label{sec:massdistance}}

The first point is that when the masses and distances of dark isolated
objects can be ``routinely'' measured, the utility of partial information
(e.g., $\theta_\e$-only measurements) will drastically decline.  In this
context, it is important to note that the technical basis for routine
BH mass-distance measurements already exists.  This may seem obvious
from the fact that there has already been one such measurement
(OGLE-2011-BLG-0462, \citealt{ob110462a,ob110462b,ob110462c}),
which had a very respectable
error of just 10\%.  However, the characteristics of this BH were
extraordinarily favorable, so that the rate of comparable-quality
BH mass measurements via the same technical path (annual parallax plus
astrometric microlensing) is likely to be very low.  First, OGLE-2011-BLG-0462
is unusually nearby\footnote{Note that in their abstract, \citet{ob110462c}
propagate the incorrect distance estimate of \citet{ob110462a}, but they
correct this in their penultimate paragraph.}
($D_L = 1.6\,\kpc$, $\pi_\rel=0.50\,\mas$),
which led to an unusually large Einstein radius
($\theta_\e=\sqrt{\kappa M\pi_\rel}=5.7\pm 0.4\,\mas$)
and (for a BH) unusually large microlens parallax
($\pi_\e=\sqrt{\pi_\rel/\kappa M}=0.088\pm 0.008$), i.e.,
both $\propto \sqrt{\pi_\rel}$.  Second, the errors in the
parallax measurement, which to leading order do not depend on the measured
values, were unusually small for two reasons.  First, while BH events
are drawn from $\sim 100\,{\rm deg}^2$ of microlensing surveys,
OGLE-2012-BLG-0462 happened to lie in the $\sim 4\,{\rm deg}^2$ of the OGLE
survey that was monitored at a high rate, $\Gamma=3\,{\rm hr}^{-1}$.  The
next highest cadence ($1\,{\rm hr}^{-1}$) would have led to errors that
would have been 1.7 times higher.  Second, being nearby (so large $\theta_\e$),
but having a typical relative proper motion $(\mu_\rel = 4.3\,\masyr)$,
meant that the event was longer than one at a typical distance for
a disk BH ($\pi_\rel\sim 60\,\muas$), by a factor $\eta\sim 2.9$.  Such
a shorter event would have had a larger error by $\eta^2\sim 8.3$ for
$\pi_{\e,E}$ (and larger for $\pi_{\e,N}$), while (as just mentioned),
$\pi_\e$ itself would be a factor $\eta$ smaller.  Hence, this distance
effect, by itself would increase the fractional error in $\pi_\e$ by
$\ga \eta^3$.  That is, for the example given, the fractional error would
be increased by a factor of 24 from 9\% to 215\%.

The relative rarity of BH events with robustly measurable $\pi_\e$
(in current experiments) interacts with the challenges of astrometric
microlensing.  In the case of OGLE-2011-BLG-0462, this required 8 years
of monitoring by {\it HST}.  If the fraction
of BH events with measurable $\pi_\e$ is small, then application of this
laborious {\it HST}-based technique cannot yield ``routine'' measurements.

This problem has already been partially solved by the development of
GRAVITY-Wide VLTI interferometry, and will be further ameliorated when
GRAVITY-Plus comes online.  GRAVITY itself can make very precise
$(\sigma \sim 10\,\muas)$ measurements \citep{kojima1}
for Einstein radii as small as $\theta_\e\ga 0.5\,\mas$.  The current
and in-progress upgrades to GRAVITY do not improve this precision (which
is already far better than required for this application), but they
permit the observation of much fainter targets.  In addition, results
can be obtained from a single observation (or two observations).  Moreover,
interferometry has a little recognized, but fundamental advantage over
astrometric microlensing: by separately resolving the images, it precisely
measures $\bmu_\rel$ including its direction.  In the great majority of
cases (although not OGLE-2011-BLG-0462), the light-curve based $\bpi_\e$
measurements yield an effectively 1-D parallax \citep{gmb94}, with errors
that are of order 5--15 times larger in one direction than the other.
By measuring the direction of lens-source motion, interferometry effectively
reduces the error in the amplitude of the parallax, $\pi_\e$, from that of
the larger component to that of the smaller component \citep{ob03175,kojima1c}.
These advances in interferometry not only greatly increase the number of
potential targets but also substantially ameliorate the difficulty of
obtaining precise parallax measurements.

Nevertheless, to obtain truly ``routine'' isolated-BH mass-distance
measurements using this approach
would require a dedicated parallax satellite in solar orbit,
which could complement ``routine'' VLTI GRAVITY high-precision $\theta_\e$
measurements, with ``routine'' satellite-parallax high-precision $\pi_\e$
measurements.  While there are draft proposals for such a satellite, there
are no mature plans.

Another path of ``routine'' BH mass measurements may open up with the
launch of the {\it Roman} space telescope.  \citet{gouldyee14} argued
that space-based microlensing observations alone could, in principle,
return mass measurements for a substantial fraction of lenses by a
combination of astrometric and photometric microlensing.  Regarding
dark objects, they explicitly excluded BDs and FFPs as unmeasurable.
Hence, here we focus only on BHs.  In this context, we note that
\citet{lam20} predicted that {\it Roman} ``will yield individual
masses of {\cal O}(100--1000) BHs''.  We briefly show that the logic
of both of these papers regarding {\it Roman} BH mass measurements is
incorrect, and that, in particular, {\it Roman} will be mostly
insensitive to bulge BHs.  Nevertheless, {\it Roman} could return
masses for some disk BHs, although, as we will show below, this issue
should be more thoroughly investigated by explicit calculations.

\citet{gouldyee14} argued that because mass determinations require both
$\btheta_\e\equiv \bmu_\rel t_\e$ and $\pi_{\e,\parallel}$
measurements, and because the former are generally substantially more
difficult (assuming they are obtained via astrometric microlensing),
one should just focus on the $\btheta_\e$ measurement when assessing
whether masses can be measured.  However, the relative difficulty is,
in fact, mass dependent, and while their assessment is valid for
typical events with $M\sim 0.5\,M_\odot$, it does not apply to BHs,
with $M\sim 10\,M_\odot$.  In particular, while the ratio of errors
$\sigma(\theta_\e)/\sigma(\pi_{\e,\parallel})$ essentially depends
only on the observational conditions, the ratio of values scales
directly with mass $\theta_\e/\pi_\e = \kappa M$.  Hence, the entire
logic of the \citet{gouldyee14} approach does not apply to BHs.
Regarding the \citet{lam20} estimate, it is rooted in very generous
assumptions, as codified in their Table~4, and it explicitly ignores the
large gaps in the {\it Roman} data stream.  In particular, they assume
that all BHs with timescales $90<t_\e/{\rm day}<300$, impact parameters
$u_0<1.7$ and source fluxes $H_{\rm AB}<26$ ($H_{\rm Vega}<24.6$)
will yield mass measurements.

While a thorough investigation of {\it Roman} sensitivity of BHs is
beyond the scope of the present work, we have carried out a few
calculations, both to check the general feasibility of this approach
and (hopefully) to motivate a more systematic investigation.  We
modeled {\it Roman} observations as taking place in two 72-day
intervals, each centered on the equinoxes of a given year, and in
three separate years that are successively offset by two years.  We
model the photometric and astrometric errors as scaling as $A^{-1/2}$
because the faint sources for which this approximation does not apply
are well below the threshold of reliable parallax measurements.  We
adopted timescales of $t_\e=60\,$day, $t_\e=120\,$day, and $t_\e=180\,$day
as representative of bulge BHs, typical-disk BHs, and nearby-disk BHs,
respectively, and we considered
events peaking at various times relative to the equinox and at various
impact parameters.  Given our assumptions, the errors in both
$\btheta_\e$ and $\pi_{\e,\parallel}$ scale inversely as the
square-root of source flux.  For purposes of discussion we reference
these results to sources with 1\% errors, i.e., $H_{\rm Vega}=20.4$,
or roughly $M_H\sim 5.3$, i.e., M0 dwarfs.

For $t_\e=60\,$day, we find that
$\sigma(\pi_{\e,\perp})/\sigma(\pi_{\e,\parallel})\sim$ 10--20.
Hence, essentially all the parallax information is in
$\pi_{\e,\parallel}$.  Because the orientations are random, this
means that one should aim for $\pi_\e/\sigma(\pi_{\e,\parallel})\sim
10$ to obtain a useful mass measurement.  For bulge BHs, i.e.,
$\pi_\rel\sim 16\,\muas$ and $M\sim 10\,M_\odot$, we expect
$\pi_\e\sim 0.014$, implying a need for $\sigma(\pi_{\e,\parallel})\la
0.0014$.  We find that this can be achieved for our fiducial sources
only for $u_0\la 0.4$ and only for offsets from the equinox of about 0
to 36 days in the direction of summer, i.e., after the vernal equinox
or before the autumnal equinox.

For typical-disk BHs, i.e., $\pi_\rel\sim 50\,\muas$ and $t_\e\sim 120\,$day,
$\pi_\e$ is larger by a factor 1.75, implying that parallax errors that
are 1.75 times larger are acceptable.  Moreover, the longer timescales
imply that the parallax measurements will be more precise.  We find that
even choosing sources that are 1.2 mag fainter (so 1.75 times larger
photometric errors), the range of allowed $t_0$ more than doubles
to the entire interval from vernal to autumnal equinox, while the range
of acceptable impact parameters increases to $u_0\la 0.6$.  The combined
effect of these changes roughly increases the fraction of events with
measurable masses by a factor $\sim 6$.  We find qualitatively similar
further improvements for near-disk lenses with $t_\e=180\,$day.

Because of the restriction to relatively bright sources,
the relatively small area covered by {\it Roman}, as well as the
limited range of allowed $t_0$ and $u_0$, mass measurements of bulge BHs
will be rare.  The situation is substantially more favorable
for disk BHs, but the restrictions remain relatively severe.
We note that we find that whenever
$\pi_{\e,\parallel}$ is adequately measured, the nominal SNR for
$\theta_\e$ is much higher.  However, we caution the reader to review
the extensive discussion by \citet{gouldyee14} of the ``known known'',
``known unknown'', and ``unknown unknown'' systematic errors.  

The prospect for mass-distance measurements of isolated substellar
objects are significantly dimmer than for isolated BHs.  Regarding FFPs,
there are proposals for new missions that would yield such measurements,
but none approved so far.  Regarding isolated BDs, there are not even
any proposals.

The only realistic way to measure $\theta_\e$ for substellar objects
is from finite-source effects.  That is, the relevant values,
$\theta_\e\la 50\,\muas$ are at least an order of magnitude smaller
than is feasible with VLTI GRAVITY and even less accessible to astrometric
microlensing using current, or currently conceived, instruments.
The event timescales are too short by 1--2 orders of magnitude to yield $\pi_\e$
from annual parallax. Hence, they must be observed from two locations, i.e.,
two locations on Earth (terrestrial parallax), or from one or several
observatories in space.  \citet{gould13} have already shown that the first
approach can yield at most a few isolated-BD mass measurements per century.

Hence, the requirement for a mass-distance measurement is that the two
observers should be separated by some projected distance $D_\perp$ that
is substantially greater than an Earth radius and that they should
simultaneously observe an event that is FSPL as seen from at least
one of them.  As a practical matter, this means that both observers
would have to be conducting continuous surveys of the same field.  The
alternative would be to alert the second observatory prior to peak,
based on observations from the first.  Because the events have timescales
$t_\e\la 2\,$days, and given constraints on spacecraft operations, there
are no prospects (also no plans) for such a rapid response
at optical/infrared wavelengths at the present
time\footnote{Such rapid response times are certainly feasible.
For example, ULTRASAT, which will observe at 230--290 nm from
geosynchronous orbit, will have a maximum response time of
15 minutes \citep{ultrasat}.  The same capacity could, in principle,
be given to microlensing parallax satellites in, e.g., L2 orbits.}.

We now assess the constraints on $D_\perp$ to make such a measurement
for substellar objects, from the standpoint of mission design.  That is,
we are not attempting to make detailed sensitivity estimates, but rather
to determine how the regions of strong sensitivity depend on $D_\perp$.
There are three criteria for good sensitivity, which we express in
terms of the projected Einstein radius, $\tilde r_\e\equiv \au/\pi_\e$,
\begin{enumerate}
\item $D_\perp \la \max(1,\rho)\tilde r_\e$.
\item $D_\perp \ga 0.05\,\max(1,\rho)\tilde r_\e$.
\item $\rho \la 3$ (for dwarfs) or $\rho \la 5$ (for giants).
\end{enumerate}

The first criterion is that the second observer will see an event, i.e.,
that the lens will pass within the maximum of
$\sim \theta_\e$ and $\sim\theta_*$ of the source on the source plane, which
translates to condition (1) on the observer plane.  While there are special
geometries for which this condition is violated and both observatories
will still see an event, our objective here is to define generic criteria,
not to cover all cases.

The second criterion ensures that the event looks sufficiently different
from the two observatories that a reliable parallax measurement
(in practice, $\ga 3\,\sigma$) can be made.  Note that \citet{gould13}
set this limit at 2\% (rather than 5\%) of the
source radius for terrestrial parallax.  However, they were considering
the case of very high cadence followup observations of very highly
magnified sources, albeit on amateur-class telescopes, whereas for the
survey case that we are considering, there are likely to be only a handful
of observations over peak.

The third criterion ensures that the event is sufficiently magnified over
peak for a reliable measurement.

\begin{figure}
\plotone{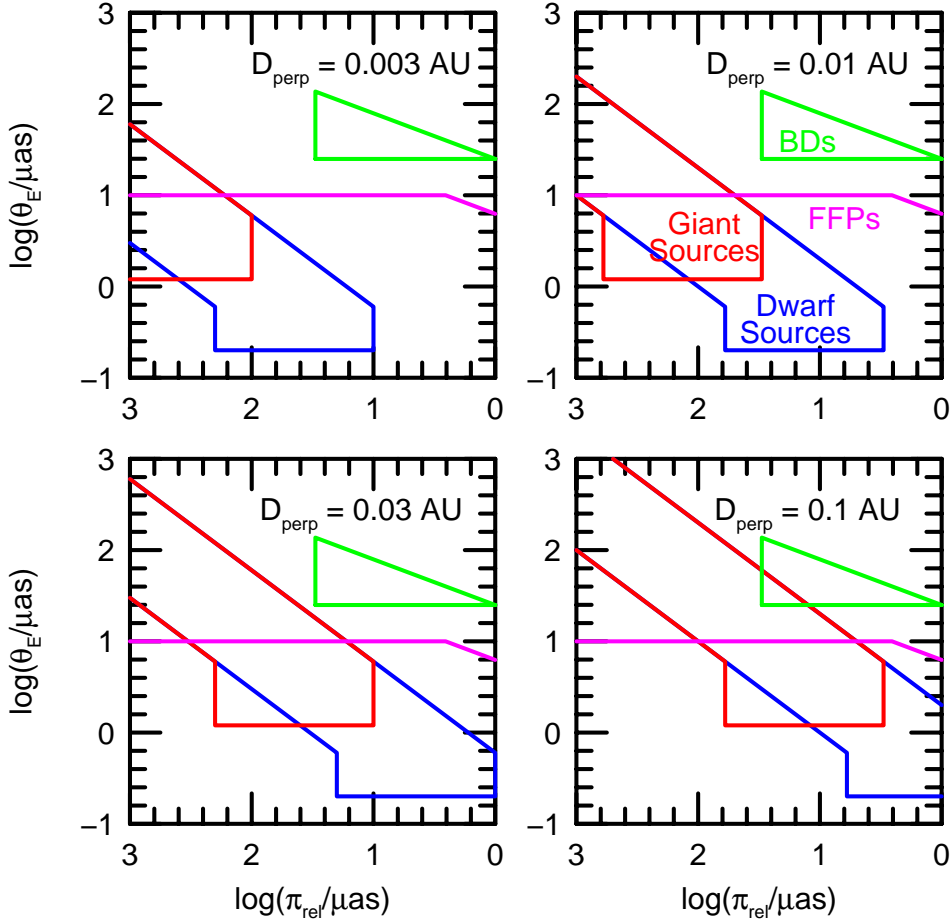}
\caption{Sensitivity to microlens-parallax measurements of substellar objects
  (BDs and FFPs) as a function of Earth-satellite separation $D_\perp$, as
  specified in the four legends.  Proposed experiments all have
  $D_\perp\sim 0.01\,\au$ (upper left panel).  The BDs (green) and FFPs
  (magenta) are separated by the observed Einstein Desert.
  For $D_\perp\sim 0.01\,\au$, there is good sensitivity to FFPs, but essentially
  no sensitivity to BDs, which would require $D_\perp\ga 0.3\,\au$.
}
\label{fig:dperp}
\end{figure}

Figure~\ref{fig:dperp} compares these constraints to the expected locations
of the two targeted populations, i.e., FFPs (magenta) and BDs (green),
for four values of $D_\perp = (0.003,0.01,0.03,0.1)\,\au$.  The axes
($\theta_\e$ versus $\pi_\rel$) are chosen to highlight what is known
about these two populations, with the central fact being that the FFPs
lie below the Einstein Desert ($\theta_\e\la 10\,\muas$), whereas the BDs
lie above it ($\theta_\e\ga 25\,\muas$).  As discussed by \citet{gould22},
there are strong reasons for believing that the BDs are in the bulge,
which we have represented by a cutoff at $\pi_\rel = 0.03\,\mas$.  We
have also imposed a somewhat arbitrary mass limit on the FFPs of
$M<5\,M_{\rm Jup}$ in order to illustrate that only if these objects are
fairly massive can they actually be in the bulge.  To illustrate the
role of dwarf and giant sources, we choose $\theta_*=0.6\,\muas$ and
$\theta_*=6\,\muas$, respectively.

Before discussing the implications of Figure~\ref{fig:dperp}, we note
that the basic form of the allowed region is a band defined by a
constant range of $\pi_\e$, i.e., $0.05 < D_\perp\pi_\e/\au < 1$, with
a somewhat complex threshold at $\pi_\rel \ga \theta_*$.

All current ideas for making these measurements are close to the top-right
panel, i.e., the Earth-L2 distance.  These include placing a satellite at
L2 to continuously observe one KMT field \citep{gould21,earth2.0},
making observations from Earth
of the {\it Roman} fields, observing the same fields simultaneously
from {\it Roman} and {\it Euclid}, both in L2 halo orbits, or observing
the same field from {\it Roman} in L2 and {\it CSST} in low-Earth orbit.
The third would be intermediate between the top-left and top-right panels.
See \citet{gould21}.  From Figure~\ref{fig:dperp}, these proposed
experiments would be well-matched to measuring FFP masses for the
known population, including members of both the disk and the bulge.
However, these experiments will not measure masses for the known BD population.
This would require $D_\perp\ga 0.3\,\au$, i.e., more in the range of what
is needed for BH mass measurements.

In summary, while some experiments proposed for the coming decade
could lead to FFP mass measurements, there are no current prospects
for BD mass measurements.  Hence, $\theta_\e$-only surveys will remain
the only method for detailed probing of this population for at least
several decades.

\subsubsection{{Prospects for FSPL Measurements}
  \label{sec:fspl}}

Another possibility is that competing approaches might obtain a
much larger number of FSPL measurements on one decade timescales
compared to what can be achieved with current experiments.  This would
diminish, although it would not negate, the urgency of making such
measurements based on current experiments.  

The main competing approach would come from the {\it Roman} telescope,
which is currently scheduled for launch in 2027 and would conduct
a total of $\sim 1.2$ years of observations of $\sim 2\,{\rm deg}^2$ at a
cadence of $\Gamma\sim 4\,{\rm hr}^{-1}$, using a broad $H$-band filter
on a 2.4m telescope at L2.

\citet{johnson20} have comprehensively studied FFP detections by {\it Roman},
including detailed attention to finite-source effects, which yield
FSPL events.  They do not extend their mass range up to BDs, but it is not
difficult to extrapolate from their maximum of $10^3\,M_\oplus$ to the BD
range.  However, our interest here is not so much the {\it absolute} number
of such detections under various assumptions, but the {\it relative} number 
compared to current experiments.

In this section, we show that while {\it Roman} will greatly increase the
number of FFP and isolated BD PSPL (i.e., $t_\e$-only) events relative to
what can be achieved from the ground, it will not be competitive in
identifying FSPL events (i.e., with $\theta_\e$ measurements), except
in the regime $0.1\la\theta_\e/\muas\la 1$.  We anticipate that the combination
of the large space-based PSPL sample with the smaller ground-based
FSPL sample will be more informative than either sample separately.
However, we do not explore that aspect here because our primary concern
is to investigate the uniqueness of the ground-based sample.

We begin by developing a new metric by which to compare the sensitivity
of the KMT and {\it Roman} samples: S/N as a function rank ordered
source luminosity.  We focus on the KMT prime fields ($\sim 13\,{\rm deg}^2$),
which have cadences similar to that of the {\it Roman} fields, i.e.,
$\Gamma=4\,{\rm hr}^{-1}$.  We will show that about 11 times more
microlensing events take place in these fields during 10 years of KMT
observations than take place in the {\it Roman} fields during its observations.
Here, we are not yet considering which of these events are actually detected
by either project.  Moreover, we are not yet restricting consideration to
FSPL events.

In this context, if we wish to compare performance on an
event-by-event basis, the events should first be rank-ordered by source
luminosity (which is the most important factor in S/N).  So, for example,
we will show that KMT sources with $M_I=3.2$ should be compared to {\it Roman}
sources with $M_I=6$, because there are the same number of microlensing
events down to these two thresholds for the respective surveys.

Because the lens-source kinematics of the two experiments are
essentially the same, the ratio of the number of events is given by
the ratio of the products $\Omega\times\Delta t \times N_S\times N_L$, where
\begin{enumerate}
  \item $\Omega$. Area of survey: $(13\,{\rm deg^2}/2\,{\rm deg}^2=6.5)$
  \item $\Delta t$. Duration of survey: $(10\times 4\,{\rm months}/6\times 72\,{\rm days} = 2.8)$
  \item $N_S$. Surface density of sources: 1/1.29.
  \item $N_L$. Surface density of lenses: 1/1.29. 
\end{enumerate}
That is, an overall ratio KMT:{\rm Roman} of $6.5\times 2.8 /1.29^2 = 11$.
Here, we have adopted field sizes of $13\,{\rm deg}^2$ for the KMT prime fields
versus $2\,{\rm deg}^2$ for {\it Roman} and durations of 4 months per year
for 10 years for KMT versus six 72-day campaigns for {\it Roman}.
In particular, we note that the KMT survey is nominally carried out
for 8 months per year, but in the wings of the season, there are huge
gaps due to the restricted times that the bulge can be observed.  Moreover,
KMT is affected by weather and other conditions (such as Moon) that
restrict the period of useful observations.  Thus, we consider that the
effective observations are 4 months per year.

%111285 roman
%BLG01  17:54:24  -31:08:00  1 -0.91 -2.76  82296
%BLG02  17:54:24  -29:01:30  1 +0.91 -1.69 107010
%BLG03  18:02:00  -27:56:30  1 +2.68 -2.60  68626 average 86000 ratio: 1.29

The mean latitude of the KMT prime fields is about
$\langle|b|\rangle_{\rm KMT}\sim 2.35^\circ$, compared to
$\langle|b|\rangle_{Roman}\sim 1.7^\circ$ for {\it Roman}.
According to \citet{nataf13}, this gives a factor of 1.29 advantage to
{\it Roman} in bulge source density.  For lenses that are in the bulge (as BDs
are expected to mainly be), the advantage is identical.  For disk lenses,
it is roughly similar.

\begin{figure}
\plotone{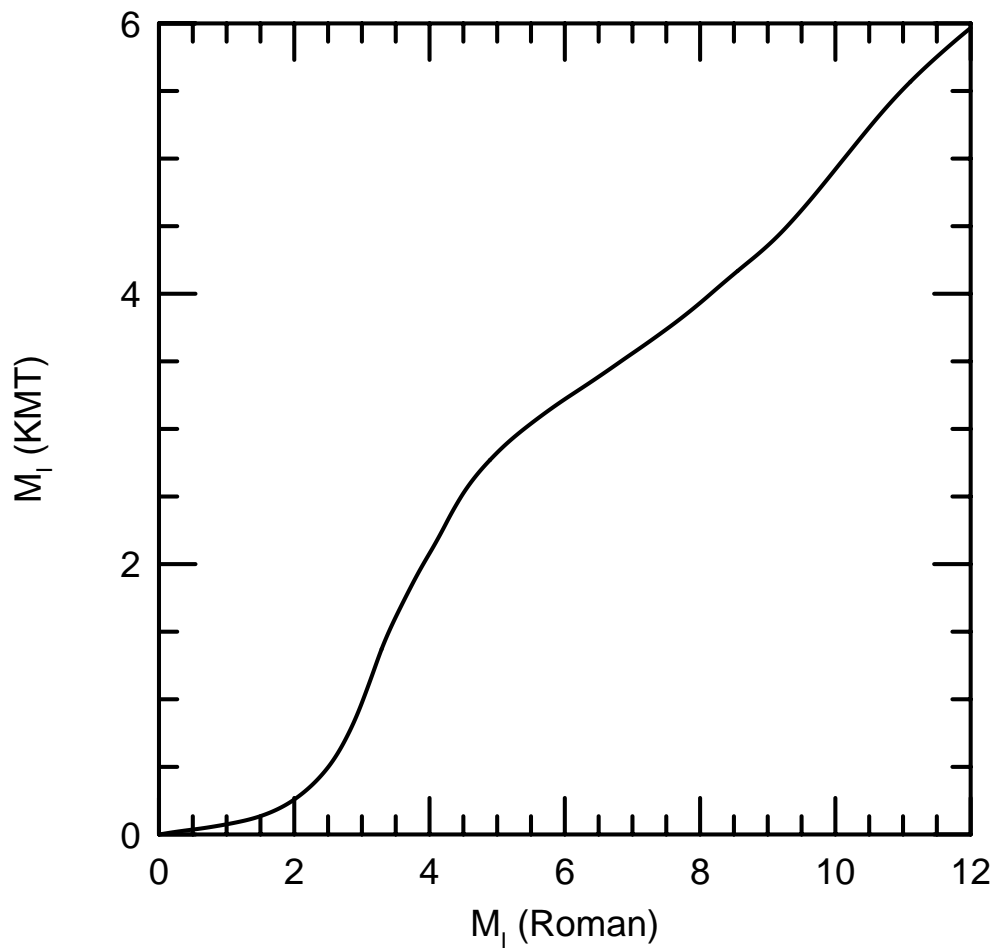}
\caption{Matched cumulative distributions of microlensing events
  taking place within the purview of the 10-year KMT survey and the
  5-year {\it Roman} survey, according to the $I$-band luminosities
  of each event.  Because 11 times more microlensing events take place
  (not necessarily detected) for KMT, relatively brighter KMT sources
  are matched one-for-one to relatively fainter {\it Roman} sources.
  As shown by Figure~\ref{fig:snrrat}, this effect tends to mitigate the
  otherwise overwhelming advantage of {\it Roman}'s higher throughput,
  lower extinction, and lower background.
}
\label{fig:matchcum}
\end{figure}

We use the HLF to calculate the cumulative distribution of sources.
This distribution is given only for $M_I\leq 9$.  However, we extend it
to $M_I=12$ (i.e., to masses $M\sim 0.1\,M_\odot$) using the \citet{chabrier05}
mass function.  We transform from mass to $I$-band luminosity using the
$V$ and $K$ mass luminosity relations of \citet{benedict16} and the color-color
relations of \citet{bb88}.  We refer to this combined luminosity function as the
CHLF.  We then adopt this as the unnormalized
{\it Roman} distribution and multiply it by 11 to construct the
KMT distribution before matching the two distributions.
Figure~\ref{fig:matchcum} shows the result.
As anticipated above, $M_I=6.00$ {\it Roman}
sources are matched to $M_I= 3.22$ KMT sources.  Note that the viability
of this approach depends on the fact that there are few useful sources
outside the diagram.  For {\it Roman} this is not an issue because the
diagram goes almost to the bottom of the main sequence.
If the matched KMT luminosity had been, say $M_{I,{\rm KMT}}=3.5$ at this point,
then the diagram would be excluding many useful KMT sources.  However,
in fact, at the actual value ($M_I=6.0$) and for most applications,
few such useful sources are being ignored.  Nevertheless, these dim KMT
sources can be important in some cases, so this must always be checked when
applying this method of matched cumulative distributions.

We now evaluate the ratio between the {\it Roman} S/N and KMT S/N at
each pair of matched values under the assumption that the source star is
unblended and at various magnifications $A$.  In each case, we assume
that what is being measured is some small change in magnification
$\Delta A$, so that the S/N ratios are
\begin{equation}
  ({\rm S/N})_X = {f_X\Delta A\over \sqrt{f_X A + B_X}}
  \label{eqn:snr}
\end{equation}
where the $f_X$ ($X=H$ or $X=I$) are the respective source flux of the matched
sources, for {\it Roman} and KMT, and the $B_X$ are the respective backgrounds.
Thus, their ratio,
\begin{equation}
  {({\rm S/N})_H\over ({\rm S/N})_I}
    = {f_H/f_I\over \sqrt{(f_H A + B_H)/(f_I A + B_I)}},
  \label{eqn:snr2}
\end{equation}
is independent of $\Delta A$.  To make these evaluations, we adopt
the following assumptions.  Regarding KMT, we assume a zero point of
1 photon per (60 second) exposure at $I_{\rm zero}=28.0$ and with a background 
$I_{\rm back}=16.8$ as described above.  Regarding {\it Roman}, we assume
a zero point of 1 photon per (52 second) exposure at
$H_{\rm zero} = 30.4$ on the Vega system and a background of $H_{\rm back}=21.7$
per exposure.  See \citet{gould14}.  These backgrounds imply
$B_I=3.02\times 10^4$ and $B_H=3.02\times 10^3$.

Next, we assume that the sources lie at $D_S=8\,\kpc$ and we adopt
an $I$-band extinction $A_I=2$, corresponding to $A_H=0.23\,A_I = 0.46$.

Finally, to convert from the $M_I$ (of the CHLF) to the required $M_H$
(needed to calculate the {\it Roman} source flux), we proceed as follows.
For M dwarf sources ($M_I\geq 6.8$),
we use the empirically calibrated mass-luminosity
relations of \citet{benedict16} in $V$ and $K$, i.e., their Equation~(10)
and Table~12.  We convert from $(V-K)$ to $(I-H)$ using the
$VIHK$ relation of \citet{bb88} and then evaluate $M_H = M_I + (I-H)$.
For the remainder of the CHLF, we use the following approximations
$(I-H) = 1.29-0.035\,M_I$, ($0\leq M_I< 2$),
$(I-H) = 1.22-0.245\,(M_I-2)$, ($2\leq M_I< 4$),
$(I-H) = 0.73\,M_I$, ($4\leq M_I< 4.5$), and
$(I-H) = 0.73+0.402\,(M_I-4.5)$, ($4.5\leq M_I< 6.8$).

\begin{figure}
\plotone{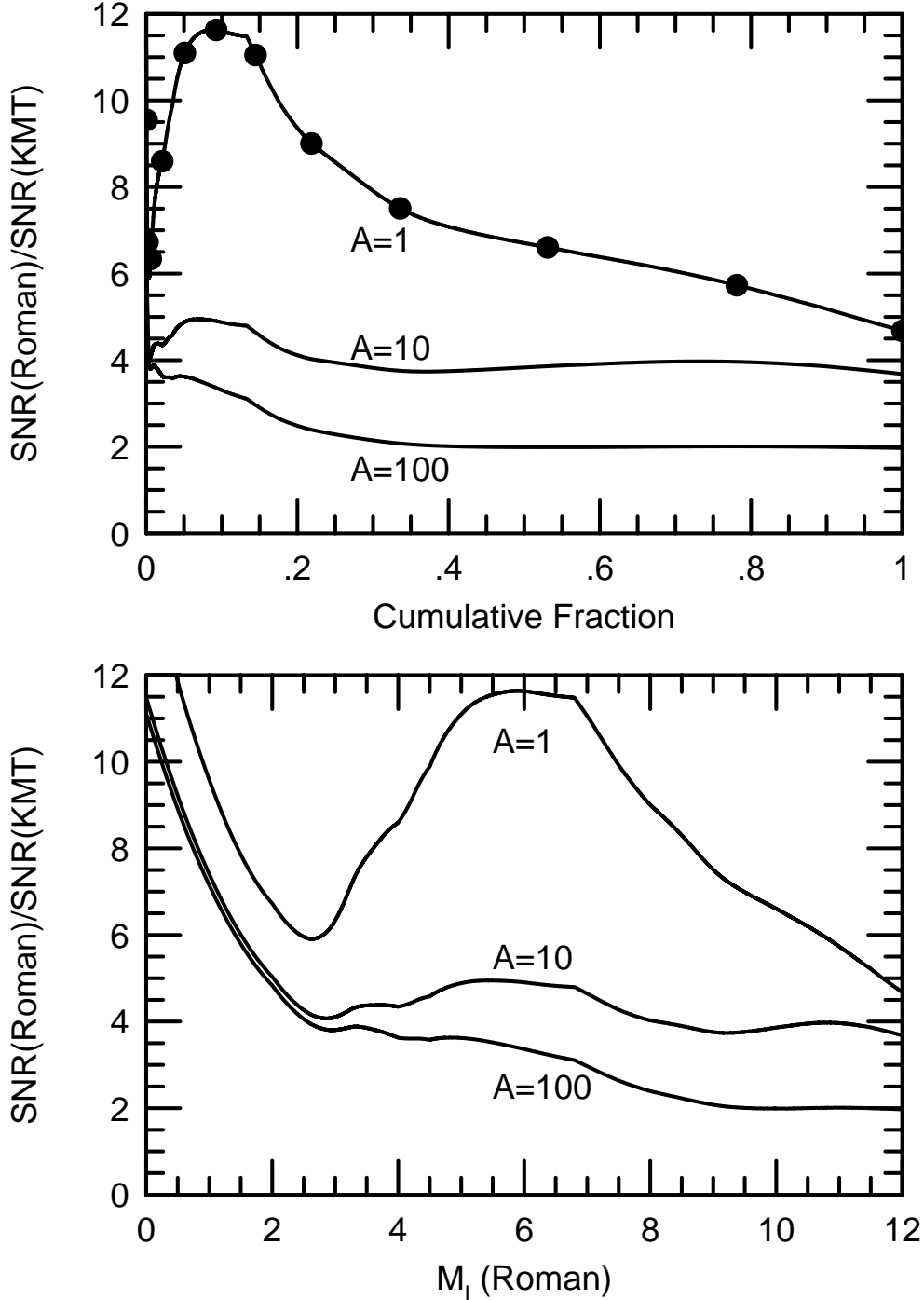}
\caption{Ratios of S/N of matched events (see
  Figure~\ref{fig:matchcum}) of {\it Roman} relative to KMT
  for three different magnifications, $A=(1,10,100)$.
  The lower panel shows these as a function of the $M_{I,Roman}$
  parameter in order to make contact with Figure~\ref{fig:matchcum}.
  Note, however, that this is a label used for matching: the {\it Roman}
  S/N is calculated in $H$-band.  The upper panel shows the same ratios
  as a function of the cumulative distribution of sources, so that
  the occurrence of microlensing events (not necessarily their detection)
  is uniformly distributed along the abscissa.  At moderate to
  high magnifications, which account for most ground-based planets,
  {\it Roman} has an advantage of a factor 2--4, whereas at low magnifications,
  {\it Roman} has an overwhelming advantage over most of this range.
  Filled circles show (right to left) $M_{I,Roman} = (12,11,\ldots)$.
}
\label{fig:snrrat}
\end{figure}

The results are shown in Figure~\ref{fig:snrrat} for three case: $A=(1,10,100)$.
The lower panel shows the S/N ratios as a function of $M_{I,Roman}$ so they can
be referenced to Figure~\ref{fig:matchcum}.  However, the implications
are best understood from the upper panel, which shows these ratios
as a function of the cumulative distribution.  The filled circles along the
$(A=1)$ curve allow one to relate the two panels.  These indicate (from right
to left) $M_{I,Roman}=(12, 11, 10, \ldots)$.  Figure~\ref{fig:snrrat} also has
important implications for the {\it Roman} bound-planet``discovery space''.
%which we discuss below.
However, the main focus of the present work
is on isolated substellar objects.  Because these objects are themselves dark,
and because they do not have a host, the assumption of ``no blending'' will
usually be satisfied.

Figure~\ref{fig:snrrat} shows that over the entire CHLF and at all
magnifications, {\it Roman} has higher S/N than KMT, implying that at each
matched luminosity, {\it Roman} will detect at least as many PSPL isolated
sub-stellar events as KMT.  Hence, it will also detect at least as many from the
CHLF as a whole.  Because {\it Roman} S/N superiority is substantial,
especially for $A=1$, over a substantial fraction of the CHLF, it may
appear that it would detect many times more PSPL events.  However,
this proves to be the case only for low-mass FFPs, whereas the factor
is more modest for isolated BDs.

The fundamental reason is that reliable detections can only be made
up to some limit, e.g., $u_0<1$, regardless of S/N.  For illustration,
we assume that such a detection is possible for KMT assuming that the
peak difference flux obeys
$\Gamma u_0 t_\e({\rm S/N})^2_{\rm peak}> \chi^2_\min=2000$.
Because KMT is background-limited, this can be written
$f_I > \sqrt{\chi^2_\min B_I/\Gamma u_0t_\e}/(A_\max-1)
\rightarrow 2300(t_\e/{\rm day})^{-1/2}$.  According to our assumption, $A_I=2$,
this corresponds to $M_I < 3.0 + 1.25\log(t_\e/{\rm day})$.  Hence, adopting
$\theta_\e=40\,\muas$ for a typical BD and assuming $\mu_\rel=6\,\masyr$
(so, $t_\e=2.4\,$day),
this implies $M_{I,{\rm KMT}} < 3.5$, which matches to $M_{I,Roman} < 6.8$
according to Figure~\ref{fig:matchcum}.  Carrying out a similar calculation
for the {\it Roman} threshold, and noting that in the relevant range it
is also background dominated, this yields
$f_H > \sqrt{\chi^2_\min B_H/\Gamma u_0t_\e}/(A_\max-1)
\rightarrow 735(t_\e/{\rm day})^{-1/2}$, i.e.,
$M_H< 8.3+ 1.25\log(t_\e/{\rm day})$.  Thus, for BDs, $M_H<8.8$, i.e.,
$M_I<11.1$.  From Figure~\ref{fig:matchcum}, {\it Roman} complete sensitivity
to these BD PSPL events covers 4.7 times more of the cumulative fraction.
Allowing for the gradual decline of $u_{0,\max}$ for KMT for dimmer sources,
we can roughly estimate that {\it Roman} will detect 4 times more BD PSPL
events, which is a relatively modest improvement.

By contrast, for FFPs with $\theta_\e=1\,\muas$, i.e., $t_\e=0.06\,$day,
the corresponding limit for KMT would be $M_I<1.5$, for which
$\theta_*\sim 3\,\muas$, i.e., $\rho\sim 3$.  Hence, these would not be
PSPL events, but rather FSPL.  We will discuss these further below.
On the other hand, for {\it Roman}, $M_H<6.8$, i.e., $M_I<8.7$ which covers
about 1/4 of the cumulative fraction.  Hence, {\it Roman} will be vastly
more sensitive to PSPL FFPs at $\theta_*=1\,\muas$.

The method of matching cumulative source distributions cannot be used
to compare {\it Roman} and KMT FSPL substellar events.  First, the matched
sources have different $\theta_*$, which is a fundamental parameter for
FSPL events,  Second, FSPL events are among the relatively rare class of
applications for which ``unmatched'' KMT sources, i.e., $M_I>6.0$, play
a crucial role.

Instead, we compare the returns of the two experiments by first setting
a threshold $\Delta\chi^2=2000$ for each, which we approximate as
$\Delta\chi^2=N_{\rm peak}[(A_\max-1)f_X]^2/(f_X + B_X)$, where
$A_\max = \sqrt{1 + 4/\rho^2}$ and
$N_{\rm peak} = 2\Gamma\theta_*/\mu_\rel\rightarrow 11.7(\theta_*/\muas)$,
and where we have made the evaluation by adopting $\mu_\rel=6\,\masyr$.
We further require $N_{\rm peak}\geq 3$ to ensure that the finite-source
effects are adequately characterized.  Finally, we demand $A_\max>1.06$,
i.e., $\rho<5.7$ because of the difficulty of distinguishing lower-amplitude
events from giant-source variability.
While these prescriptions are
simplified, they are adequate to characterize the relative sensitivity
of the two experiments.  For the source radii, we adopt
$\theta_* = 6\times 10^{-0.2\,M_I}\muas$, ($0\leq M_I< 2$),
$\theta_* = 10^{0.378-0.3.01(M_I-2)}\muas$, ($2\leq M_I< 4$),
$\theta_* = (R_\odot/D_S)\times 10^{-0.2\,(M_I-4.07)}$, ($4\leq M_I< 4.5$), 
$\theta_* = 10^{-0.320-0.041(M_I-4.5)}\muas$, ($4.5\leq M_I< 6.8$), and
$\theta_* = (M/M_\odot)(R_\odot/D_S)$, ($M_I> 6.8$),
where $D_S=8\,\kpc$.

\begin{figure}
\plotone{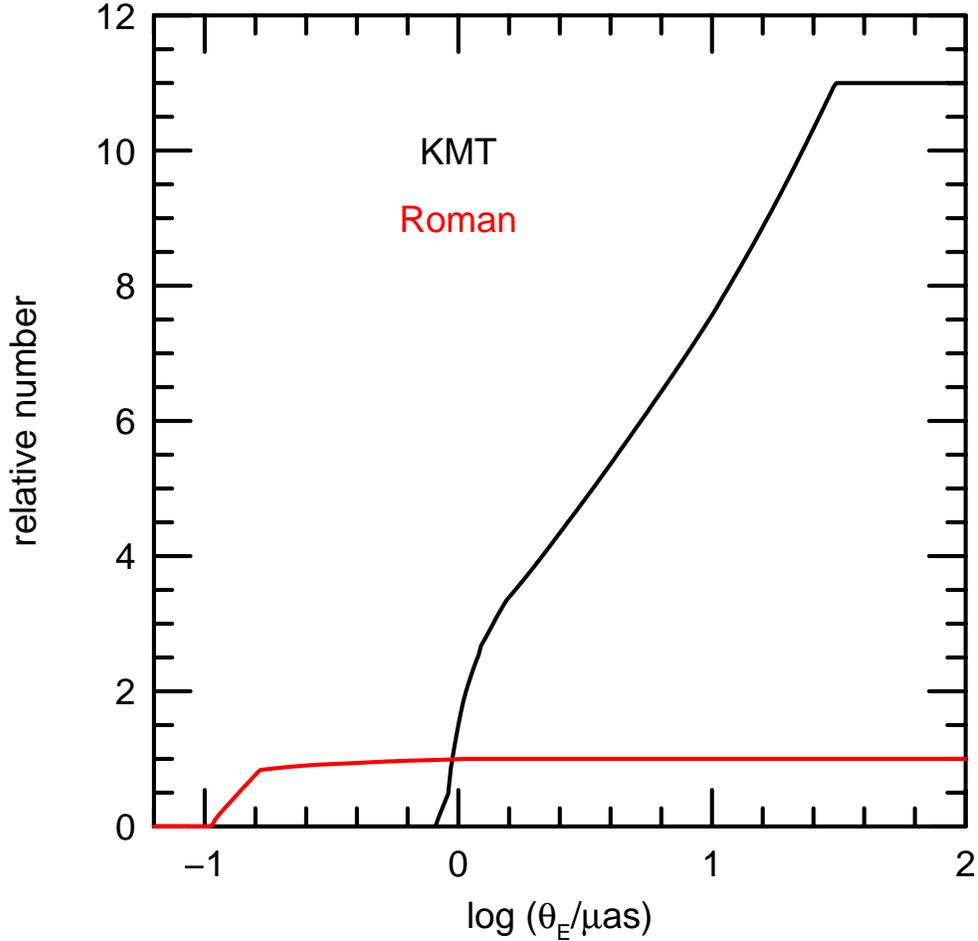}
\caption{Relative sensitivity to FSPL events of the {\it Roman} and
  KMT surveys as a function of Einstein radius $\theta_\e$.  The
  asymptotic ratio of 11 at the right reflects the fact that 11 times
  more events take place in the KMT purview, while essentially all
  FSPL events with $\theta_\e\sim 100\,\muas$ are detected in either survey.
  For smaller $\theta_\e$, {\it Roman} continues to detect almost all events
  down to almost $\theta_\e\sim 0.1\,\muas$, whereas KMT sensitivity
  continuously declines over this range, dropping to zero near
  $\theta_\e\sim 1\,\muas$.  This points to $0.1\la \theta_\e/\muas\la 1$
  as a key ``discovery space'' for {\it Roman}, a region that could
  contain a vast population of FFPs.
}
\label{fig:fspl}
\end{figure}

The results are shown in Figure~\ref{fig:fspl}.  There is a rapid
transition at $\theta_\e\sim 1\,\muas$: below this threshold, KMT
loses all sensitivity, while {\it Roman} retains constant sensitivity
for almost a decade; above the threshold, KMT completely dominates
the detections, reaching a factor 11 in the BD regime, $\theta_*\ga 30\,\muas$.
The physical reason for this dominance is simple.  At the adopted threshold
of detectability, $\theta_{*,\rm thresh}=(3/2)\mu_\rel/\Gamma=0.256\,\muas$,
i.e., $M_{I,\rm thresh} = 8.25$ or $I_S=24.75$, the source is magnified by
$A_\max = 2/\rho\rightarrow 230$ to $I\sim 18.8$, which creates a marginally
detectable event.  Thus, all sources (down to this threshold) yield detectable
events from either KMT or {\it Roman}, but because the former has 11 times
more events, it has 11 times more detections.  In fact, the completeness
analysis of Section~\ref{sec:relative} shows that only of order half of such
high-magnification events from uncataloged sources are recovered by
current KMT searches.  This could be rectified by additional specialized
searches for such ``spike events'' but even without such an effort, KMT will
still dominate in this regime.

By the same token, at $\theta_\e= 1\,\muas$, faint sources are insufficiently
magnified to boost them to detectability.  For example, for near-optimal
sources,
$\theta_*=\theta_\e$, i.e., $M_I=3.28$ or $I_S=19.78$ and with peak magnification
$A_\max = \sqrt{5}$, the difference magnitude is only $I_{\rm diff}=19.55$.
However, based on its far greater flux counts and lower background,
such events are easily detected by {\it Roman}.

We emphasize that the ``flat'' form of the {\it Roman} curve over 3 decades
does not mean that one expects equal numbers of detections across this range:
based on what we know today \citep{gould22}, there could be of order 40
times more FFPs at $\theta_* = 10^{-0.8}\muas$ compared to
$\theta_* = 10^{+0.8}\muas$.

\subsubsection{{Summary}
\label{sec:summary}}

The only way to press forward the study of isolated bulge BDs is by FSPL events
from ground-based surveys, mainly KMT.  It is not possible to obtain a
substantial number of full mass-distance measurements for these objects
from any current, planned, or proposed experiments.  Regarding FSPL BD events,
there are no other current or currently planned experiments that could
compete with KMT.

The situation is more nuanced for FFPs.  First, in the next decade, new
experiments could yield mass-distance measurements, assuming that the
current proposals for these are approved and implemented.  Second,
{\it Roman} will be increasingly competitive for FFPs within the range
$2\ga\theta_\e/\muas \ga 1$ and will be completely dominant for
$\theta_\e\la 1\,\muas$.

\subsection{{Comments on the Recent MOA FSPL Search}
\label{sec:moaffp}}

As the present paper was being completed, \citet{moaffp} reported results
from a comprehensive search for FSPL events in Microlensing Observations in
Astrophysics (MOA) Collaboration data over the 9 years from 2006 to 2014.
Here we comment on a few implications that relate to results and ideas
that we have presented.

The most important point is that \citet{moaffp} searched for FSPL events
with both giant and dwarf sources, which is what we have broadly advocated
here.  In particular, the inclusion of ``dwarf'' (including main-sequence and
subgiant) sources led to the discovery of
a very small FFP, MOA-9yr-5919, with $\theta_\e=0.90\pm 0.14\,\muas$.
Being the second such discovery (after OGLE-2016-BLG-1928,
$\theta_\e=0.84\pm0.06\,\muas$, \citealt{ob161928}), it strongly implies
that such objects are very common.  That is, a single such
discovery would be consistent with a low-probability, e.g., $p=5\%$,
detection from a relatively rare population.  However, two such
chance discoveries would occur only at ${\cal O}(p^2)$.  It was exactly
this logic that led \citet{ob05169} to conclude that
``Cool Neptune-like Planets are Common'' based on two detections, which was soon
confirmed by \citet{ob07368} and then, subsequently, by of order two dozen
other detections (\citealt{suzuki16}, Zang et al., in prep).
The inclusion of dwarf sources was crucial
to this discovery: if the same planet had transited a typical source
from the \citet{gould22} giant-star survey, with $\theta_*\sim 6\,\muas$,
it would have had $\rho\sim 6.7$ and hence excess magnification
$A-1\simeq 2/\rho^2 \sim 4.5\%$ and would not have been detected.

\citet{moaffp2} show (their Table~3), that the MOA and KMT surveys
are consistent in their constraints on the FFP population, including
both the power-law index and its normalization $Z$.
In particular, at the same zero-point, they find $Z=0.53^{+0.19}_{-0.40}$ versus
$0.39\pm 0.20$ FFPs per dex per (stars + BDs) for KMT.

At first sight, one may wonder about the consistency of the detection rates
of the KMT giant-source survey, which discovered 29 FSPL events satisfying
$I_0<16.5$, with the ``giant component'' of the MOA survey,
with 7 such events.  However, we now show that the ratio of detections
is consistent with expectations. First,  \citet{moaffp} note
that they are insensitive to the biggest $(\theta_*\ga 10\,\muas)$ sources
from the KMT survey due to saturation.  (For many of these bright sources,
KMT recovered from saturation using $V$-band observations.)\ \
From Figure~2 of \citet{moaffp2}, such a cut would eliminate $\sim 1/3$
of KMT events.  Second, the MOA detector is about half the size of the KMT
detectors (2.2 versus 4 square degrees), and in line with this fact, it
surveys roughly half the area (i.e., mainly southern bulge versus full bulge).
Third, the KMT survey employs three telescopes, whereas the MOA survey uses
one.  Moreover, while MOA and KMTA have comparable conditions,
KMTS and KMTC have better conditions.  If we were considering very short
events, which are mainly localized to a single observatory, then this
would give KMT a 4:1 advantage.  However, because giant-source events
are very long, often covering two or more observatories, we reduce
this estimate to 3:1.  Finally, the MOA survey covered 9 years while
the KMT covered 4 years.  Combining all factors, we expect a ratio
of MOA-to-KMT giant-source FSPL events of
$(2/3)\times (1/2) \times (1/3)\times (9/4) = 25\%$ compared to an
observed ratio of 24\%, which is consistent.

\acknowledgments
This research has made use of the KMTNet system operated by the Korea Astronomy and Space Science Institute (KASI)  at three host sites of CTIO in Chile, SAAO in South Africa, and SSO in Australia.  Data transfer from the host site to KASI was supported by the Korea Research Environment Open NETwork (KREONET).
Work by C.H. was supported by the grants of National Research Foundation 
of Korea (2020R1A4A2002885 and 2019R1A2C2085965).
J.C.Y., S.-J. C., and I.-G. S acknowledge support from US NSF Grant No. AST-2108414.
Y.S. acknowledges support from BSF Grant No. 2020740.
W.Z. and H.Y. acknowledge support by the National Science Foundation of China (Grant No. 12133005).

\begin{deluxetable}{llrr}                                             
\tablecolumns{4} \tablewidth{0pc}                                     
\tablecaption{\textsc{KMT-2022-BLG-2397 Light Curve Parameters}}      
\tablehead{                                                           
\colhead{Parameter} &                                                 
\colhead{Units} &                                                     
\colhead{Value} &                                                     
\colhead{Error}                                                       
}                                                                     
%\hline                                                                
\startdata                                                            
$t_0$              & (HJD$^\prime$) & 9812.31849 & 0.00035\\
$t_\e$             & (day)          &       1.35 &    0.11\\
$t_\eff$           & (day)          &    0.03588 & 0.00087\\
$t_*$              & (day)          &    0.05016 & 0.00065\\
$f_S t_\e$         & (day)          &     0.1047 &  0.0025\\
$u_0$              &                &     0.0266 &  0.0025\\
$\rho$             &                &     0.0371 &  0.0032\\
$I_{\rm S,KMTC03}$ &                &      20.78 &    0.11\\
$z_0$              &                &      0.715 &   0.017\\
%\hline                                                                
\enddata                                                              
\tablecomments{$t_\eff$, $t_*$, $f_S t_\e$, and $z_0$
are derived parameters and are not fit separately.}                   
\label{tab:parms}                                                     
\end{deluxetable}

\clearpage

\end{document}